\documentclass[12pt]{article}

\usepackage{jheppub}
\makeatletter
\def\@fpheader{\relax}
\makeatother

\usepackage{amsthm,amsmath,amssymb,mathtools}
\usepackage{mathrsfs}
\usepackage{bm}
\usepackage{bbm}
\usepackage{float}
\usepackage{graphicx}
\usepackage{caption}
\usepackage{subcaption}
\usepackage[most]{tcolorbox}
\usepackage[mathscr]{euscript}
\usepackage{dsfont}
\usepackage[mathscr]{euscript}
\usepackage{mathrsfs}
\usepackage{hyperref}
\usepackage{tikz}
\usetikzlibrary{shapes.geometric}
\usetikzlibrary{arrows.meta}
\usetikzlibrary{decorations.markings}

\newcommand\blfootnote[1]{% 
	\begingroup 
	\renewcommand\thefootnote{}\footnote{#1}% 
	\addtocounter{footnote}{-1}% 
	\endgroup 
}
\def\mN{\mathcal{N}}

\newenvironment{nohyphens}{%
	\hyphenpenalty=10000
	\exhyphenpenalty=10000
	\sloppy %
}{\par}

\title{Modularity in $d>2$ free conformal field theory}

\author[a]{Yang Lei\blfootnote{*The authors are ordered purely alphabetically and should all be viewed as the co-first authors. }}
\author[b]{\!\!, Sam van Leuven}

\affiliation[\,a]{Institute for Advanced Study \& School of Physical Science and Technology,  \\ Soochow University, Suzhou 215006, P.R.~China}
\affiliation[\,b]{Mandelstam Institute for Theoretical Physics, School of Physics, NITheCS,  and
DSI-NRF Centre of Excellence in Mathematical and Statistical Sciences (CoE-MaSS), \\ University of the Witwatersrand, Johannesburg 2050, South Africa}

\emailAdd{leiyang@suda.edu.cn}
\emailAdd{sam.vanleuven@wits.ac.za}

\abstract{\begin{nohyphens}
	We derive new closed form expressions for the partition functions of free conformally-coupled scalars on $S^{2D-1}\times S^1$ which resum the exact high-temperature expansion. The derivation relies on an identification of the partition functions, analytically continued in chemical potentials and temperature, with multiple elliptic Gamma functions. These functions satisfy interesting modular properties, which we use to arrive at our expressions. We describe a geometric interpretation of the modular properties of multiple elliptic Gamma functions in the context of superconformal field theory. Based on this, we suggest a geometric interpretation of the modular property in the context of the free scalar CFT in even dimensions and comment on extensions to odd dimensions and free fermions.
	\end{nohyphens}
}

\date{}

\begin{document} 
	
	\maketitle
	
	\section{Introduction}\label{sec:intro}

In the seminal work \cite{Cardy:1986ie}, Cardy showed that modular invariance imposes powerful constraints on the operator spectrum of two-dimensional conformal field theories.
In particular, the operator growth at large conformal dimension is completely determined by the conformal anomaly of the theory.
Many other applications followed.
One early application was towards the classification of rational conformal field theories \cite{Cappelli:1987xt} and their fusion rules \cite{Verlinde:1988sn}.
In addition, through holography, there have been important applications to quantum gravity, such as the interpretation of black hole entropy \cite{Carlip:1995qv,Strominger:1996sh,Strominger:1997eq} and the gravitational path integral \cite{Dijkgraaf:2000fq,Witten:2007kt,Maloney:2007ud}.
Holography, in turn, inspired the modular bootstrap program as formulated in \cite{Hellerman:2009bu} and the notion of a ``holographic CFT'' \cite{Hartman:2014oaa} singles out interesting classes of modular forms \cite{Belin:2016knb}.

The connection between modular invariance and the operator content is generally not expected for conformal field theories in higher dimensions.
Indeed, through the operator-state correspondence, the operator content is encoded in the partition function on $S^{d-1}\times S^1$ with no apparent modular symmetries for $d>2$.\footnote{Modular properties in higher dimensions do occur for background manifolds involving tori, but such backgrounds are not directly connected to the (local) operator content of the CFT \cite{Shaghoulian:2015kta,Shaghoulian:2015lcn,Belin:2016yll,Belin:2018jtf,Luo:2022tqy,Alessio:2021krn,aggarwal2024}.}
Somewhat surprisingly, the partition function of a free conformally coupled scalar in $d$ dimensions does admit some notion of modularity, but it is not clear to what extent such properties extend to non-trivial CFTs \cite{cardy1991operator} (see also \cite{Kutasov:2000td}).
Quite likely, this is related to the integrable properties of the free theory, reflected in its higher spin symmetries. 
In the case of higher-dimensional strongly-coupled CFTs, other successful methods have been developed to determine the asymptotic density of states, for example through holography \cite{Carlip:1998wz,Verlinde:2000wg,Cai:2001jc,Cai:2005kw,Gibbons:2005vp,Harlow:2021trr,Kang:2022orq}, and more recently making use of the framework of thermal effective field theory \cite{Benjamin:2023qsc,Allameh:2024qqp,Benjamin:2024kdg}.
Existing geometric arguments for modular symmetry in higher dimensions seem to involve a change of topology of the background \cite{Shaghoulian:2016gol,Horowitz:2017ifu}. 

For supersymmetric CFTs in higher dimensions, unconventional modular properties have recently surfaced for which a geometric interpretation akin to two dimensions seems to be possible.
For example, the works \cite{Cheng:2018vpl,Cheng:2023row,Cheng:2024vou} study modular properties of so-called homological blocks, which can be understood in terms of a supersymmetric partition function of 3d $\mathcal{N}=2$ SCFTs on a solid torus. 
Depending on the nature of the 3d bulk theory and the coupled 2d boundary theory, the homological blocks exhibit a variety of interesting modular properties. 

In four dimensions, interesting observations have been made around the supersymmetric partition function on $S^{3}\times S^1$, known as the superconformal index \cite{Romelsberger:2005eg,Kinney:2005ej,Bhattacharya:2008zy}.
This object encodes aspects of the BPS operator content of a theory.
More precisely, it counts differences between the numbers of bosonic and fermionic operators.
Crucially, it is a renormalization group invariant, which allows one to calculate this object for a rather general class of (strongly coupled) SCFTs.
The study of this object has led to Cardy-like constraints on the (BPS) operator spectrum in terms of the conformal (and 't Hooft) anomalies of 4d and 6d SCFTs \cite{DiPietro:2014bca,ArabiArdehali:2015ybk,DiPietro:2016ond}, see also \cite{Cabo-Bizet:2018ehj,Benini:2018ywd, Goldstein:2019gpz,ArabiArdehali:2019tdm,Kim:2019yrz,Cabo-Bizet:2019osg,Goldstein:2020yvj,Cassani:2021fyv} for applications to AdS$_5$ black hole entropy.

As it turns out, underlying these works there is also a notion of modularity.
A key building block of 4d superconformal indices, the elliptic Gamma function, is known to fit into a generalized modular structure based on $SL(3,\mathbb{Z})$ \cite{Felder_2000}.
A geometric explanation of this structure was pioneered in \cite{Gadde:2020bov} and extended to other dimensions as well.\footnote{Earlier work focused on the modular properties of the so-called Schur index, a specialization of the superconformal index. This started with \cite{Razamat:2012uv} and was explored more recently in \cite{Pan:2021mrw,Beem:2021zvt}.
The occurrence of modularity in this context is in part explained due to a connection of the Schur index with characters of vertex operator algebras (VOAs) \cite{Beem:2013sza,Beem:2017ooy}. A full geometric understanding, however, seems to be lacking. Similar results were also obtained for the thermal partition function of (non-supersymmetric) large $N$ Yang-Mills theory in the free limit and confined phase \cite{Basar:2015xda} (see also \cite{Basar:2015asd}).}
In \cite{Jejjala:2022lrm}, this idea was further explored.
The key idea relies on a factorization property of the index, which reflects the fact that $S^3\times S^1$ can be split into two disk-times-torus ($D_2\times T^2$) geometries with boundary $T^3$.\footnote{This follows from the genus one Heegaard splitting of $S^3$.} 
This property of the index is known as holomorphic block factorization \cite{Nieri:2015yia}, originally studied in the context of three dimensions \cite{Beem:2012mb} (holomorphic blocks are closely related to the homological blocks mentioned above, see, \textit{e.g.}, \cite{Gukov:2017kmk} for a definition of the latter).
Remarkably, when the 4d index is expressed in factorized form, it has an unconventional type of modular property with respect to a general $SL(2,\mathbb{Z})$ transformation acting simultaneously on both tori in the $D_2\times T^2$ geometries.
The $SL(2,\mathbb{Z})$ action cannot be understood in terms of large diffeomorphisms of the full background, as expected, and as such the index cannot be written in terms of ordinary modular forms.
Instead, the modular action reflects ambiguities in a choice of time circle in the split geometry.
For this reason, the generalized modular structure is referred to as \emph{modular factorization}.
See \cite{Jejjala:2022lrm} for more details.
There are strong indications that this basic idea extends to superconformal indices in other dimensions, based on known modular properties of the so-called multiple elliptic Gamma functions.\footnote{We present an overview in Section \ref{ssec:mod-mult-gamma}, including the relevant references.} 

In this work, we draw lessons from the supersymmetric story to shed new light on the modular structure underlying free, non-supersymmetric CFTs.
In particular, we follow a recent observation by Nekrasov \cite{Nekrasov:2023xzm} that the analytic continuation of parameters in a variety of physical quantities can be naturally realized in supersymmetric quantum field theory.
Inspired by this, we first show that the multiple elliptic Gamma function $G_{2D-1}(z|\underline{\tau})$ provides an analytic continuation in temperature and angular potentials of the thermal partition function of a conformally coupled free scalar on $S^{2D-1}\times S^1$. 
Using known modular properties of $G_{2D-1}(z|\underline{\tau})$, we derive exact, closed form expressions for the scalar partition functions in four and six dimensions in terms of a modular transformed argument (\textit{i.e.}, $\beta\to 4\pi^2/\beta$).
For example, when all angular potentials are turned off, we find the following expression in four dimensions:
\begin{equation}\label{eq:ex-result}
		Z_4(\tau) = \exp \left[- \frac{2\pi i}{120} B_{5}(\tau+1;\tau)\right] \frac{1}{\psi_4(3)}\prod_{k=0}^\infty \frac{\psi_4 (\frac{\tau+k+1}{\tau})}{\psi_4(4-\frac{\tau-k-1}{\tau})}\,,
\end{equation}
where $\tau=i\frac{\beta}{2\pi}$, $Z_4(\tau)$ is the 4d partition function, $B_5(z;\tau)$ is a degree five polynomial in $z$ and $\psi_4(z)$ can be expressed in terms of the exponential of four polylogarithms $\text{Li}_{r\leq 4}(e^{-2\pi i z})$.
This expression provides the exact high-temperature asymptotics $\tau\to 0^{+i}$ of the partition function, including non-perturbative corrections, as we will demonstrate.
Our expression can be viewed as a resummation of formulas recently derived in an appendix of \cite{Benjamin:2023qsc}.
Apart from this, our formula exhibits an underlying, unexpected modular structure.
We further extend these results to include angular potentials and to general modular transformations in order to study the ``root-of-unity'' asymptotics $\tau\to -\frac{n}{m}+0^{+i}$ and find consistency with the very recent work \cite{Benjamin:2024kdg}.
Finally, we suggest that \eqref{eq:ex-result} provides an example of the generalization of modularity to higher dimensions and speculate on a geometric interpretation of the formula inspired by modular factorization. 

The rest of the paper is organized as follows. 
In Section \ref{sec:review-scalar}, we review modular aspects of the free scalar partition functions in both $d=2$ and $d\ge 3$, including Meinardus' theorem.
In Section \ref{sec:modularity-higher}, we deduce a relation between the multiple elliptic Gamma functions and the free scalar partition function on $S^{2D-1}\times S^1$. 
By applying modular properties of multiple elliptic Gamma functions, we derive our expressions the partition functions and demonstrate how they encode the exact high-temperature asymptotics.
In Section \ref{sec:generalimodualr}, we extend this to general $SL(2,\mathbb{Z})$ transformations and the corresponding root-of-unity asymptotics. 
Finally, in Section \ref{sec:discussion} we speculate on a geometric interpretation of our expressions and comment on applications and generalizations of our results.

\section{Review of the free scalar partition function}\label{sec:review-scalar}

The study of the asymptotics of $d=2$ CFT partition functions was initiated by Cardy in \cite{Cardy:1986ie} and extended to $d\ge 3$ in \cite{cardy1991operator}.
In two dimensions, one makes use of modular properties of the partition function.
One method to \emph{derive} the modular property, in the context of free CFTs, is based on a reflection formula of the Riemann $\zeta$-function.
This method can be extended to higher dimensions and, even though a modular structure akin to 2d does \emph{not} arise, there are still interesting traces of modularity and, in particular, pertains to the high-temperature asymptotics of free CFT partition functions.
To contrast with our proposal in Section \ref{sec:modularity-higher}, in this section we review this methodology, with a focus on the free conformally coupled scalar on $S^{d-1} \times S^1$ for $d\geq 2$.
Our discussion follows the more recent work \cite{Melia:2020pzd}, where these techniques were revisited and applied to the asymptotic operator degeneracies in EFTs.
We also briefly comment on the work \cite{Benjamin:2023qsc}, which calculates, in an appendix, the \emph{exact} high-temperature asymptotics for the free scalar.

\subsection{Mellin representation of the partition function}\label{ssec:mellin}

The partition function of a CFT quantized on $S^{d-1}$ is given by:
\begin{equation}\label{eq:def-partitionfunction}	
Z (\beta) = \mathrm{Tr}_{\mathcal{H}(S^{d-1})}\,e^{-\beta( H-i\Omega_i J_i)}\,, 
\end{equation}
where $H$ is the Hamiltonian on $S^{d-1}$ and we have included charges for the independent angular momenta $J_i$ on $S^{d-1}$.
In the following, we will often use the notation
\begin{equation}\label{eq:beta-omega-tau-theta}
    \tau = i\frac{\beta}{2\pi } \,,\qquad \theta_i=\frac{\beta}{2\pi} \Omega_i\,.
\end{equation}
In free field theories, one can write the full ``multi-particle'' partition function $Z(\beta)$ in terms of a ``single-particle'' partition function $Y(\beta)$ via plethystic exponentiation:
\begin{equation}\label{eq:Phythestic-exp}
Z(\beta)=	\text{PE}[Y(\beta)] = \exp \left[\sum_{m=1}^\infty \frac{1}{m} Y(\beta^m) \right]\,.
\end{equation}
To derive the $\beta\to 0$ asymptotics of $Z(\beta)$, it turns out to be useful to represent it via a Mellin transformation of the so-called Hamiltonian zeta function.
The Hamiltonian $\zeta$-function is defined as (see, \textit{e.g.}, \cite{Cardy:1986ie,cardy1991operator,Gibbons:2006ij}): 
\begin{equation}\label{eq:ham-zeta}
D(d,s) = \text{Tr}(H^{-s}) = \frac{1}{\Gamma(s)} \int_0^\infty Y(\beta) \beta^{s-1} d\beta\,.
\end{equation}
where $d$ denotes the spacetime dimension. 
The partition function can now be written as the inverse Mellin transformation with in addition the insertion of a Riemann zeta function $\zeta(s+1)$ kernel:
\begin{equation}\label{eq:Mellin-rep-partition}
	\ln Z(\beta) = \frac{1}{2\pi i} \int_{\gamma-i\infty}^{\gamma+i \infty} \beta^{-s} \Gamma(s) \zeta(s+1) D(d,s)ds \,,
\end{equation}
where $\gamma$ is a positive number such that the integrand has no poles for Re$(s)>\gamma$.

Consider for example the thermal partition function of a $d$-dimensional conformally coupled free scalar  \cite{cappelli1989stress,cardy1991operator,Kutasov:2000td,Beccaria:2014jxa,Giombi:2014yra}.
The partition function with all chemical potentials turned on is given by \cite{Melia:2020pzd,Nekrasov:2023xzm}
\begin{align}\label{eq:partitionfunction-anyD}
\begin{split}
    Z_{2D}(\tau,\theta_{1},...,\theta_D) &= \exp \left[ 
\sum_{l=1}^{\infty} \frac{q^{-l}-q^l}{l} \prod_{i=1}^D \frac{1}{q^l-2\cos (2\pi l \theta_i) +q^{-l}}
\right]\,,\\ 
Z_{2D+1}(\tau,\theta_{1},...,\theta_D,\xi) &= \exp \left[ 
\sum_{l=1}^{\infty} \frac{q^{-\frac{l}{2}}+q^{\frac{l}{2}} \xi^l}{l} \prod_{i=1}^D \frac{1}{q^l-2\cos (2\pi l \theta_i) +q^{-l}}
\right]\,, 
\end{split}
\end{align}
where $q=e^{2\pi i\tau}$ and $\theta_i$, defined in \eqref{eq:beta-omega-tau-theta}, and $\xi=\pm 1$ label $O(d)$ twisted boundary conditions. 
The partition function with $\theta_i$ turned off is given by
\begin{equation}\label{eq:scalar-partition-unrefined-d}
	Z_d (\tau)= \exp \left[ \sum_{l=1}^\infty \frac{1}{l}
	\frac{q^{ \frac{d-2}{2}l } (1-q^{2l})}{(1-q^l)^d}
	\right] \,.
\end{equation}
The single-particle partition function, \textit{i.e.}, the exponent, can be interpreted as follows. 
In $d$ dimensions, the conformal dimension of the free scalar is $(d-2)/2$. 
The factor $(1-q)^d$ in the denominator reflects the $d$ derivatives acting on a single-letter operator while the factor $1-q^2$ in the numerator quotients by the equations of motion. 
Finally, the density of states is calculated through an inverse Legendre transformation 
\begin{equation}
	\rho(\Delta) = \int d\beta Z(\beta) e^{\beta \Delta} \,.
\end{equation}

Making use of the Mellin transformation, we now continue with the high-temperature asymptotics of partition function and density of states in even dimensions $d=2,4,6$.
The method also applies to odd dimensions, but we will only consider even dimensions in this work for reasons we come back to in Section \ref{sec:discussion}.

\subsection{\texorpdfstring{$d=2$}{d=2}}\label{ssec:2d}

In two dimensions, the free scalar partition function can be written as
\begin{align}\label{eq:parti-2d}
	\begin{split}
Z_{2} (\tau,\bar{\tau})&= \exp\left[
\sum_{l=1}^\infty \frac{1}{l} \left(
\frac{(1-q^l \bar{q}^l)}{(1-q^l)(1-\bar{q}^l)} -1
\right)
\right]  = Z_2(\tau) \bar{Z}_2(\bar{\tau})
	\end{split}
\end{align}
where we included a chemical potential for spatial rotations such that left- and right-movers are distinguished by $q= e^{2\pi i\tau}$ and $\bar{q}=e^{-2\pi i \bar{\tau}}$.
Furthermore, we have not included the Casimir energy for uniformity with the higher-dimensional examples.
Finally the well-known factorization into left- and right-movers is captured by
\begin{equation}
Z_2(\tau) = \exp \left[\sum_{l=1} \frac{1}{l} \frac{q^l}{1-q^l} \right] = \prod_{l=1}^\infty \frac{1} {(1-q^l)}
\end{equation}
The presence of $-1$ in the exponent of \eqref{eq:parti-2d} is specific to two dimensions \cite{Melia:2020pzd}.
It reflects the fact that 2d scalars are dimensionless and therefore there is an infinite amount of operators with the same conformal dimension obtained by decorating a given operator with a function $f(\phi)$. 
The $-1$ cures the consequent divergence and can be viewed as modding out by the equivalence relation between operators which only differ by such a function $f(\phi)$.

Up to the ignored Casimir energies, $Z_2(\tau)$ is precisely the Dedekind eta function.
Using this connection, one deduces that $Z_2(\tau)$ satisfies the following modular property
\begin{equation}\label{eq:modularity-eta}
	Z_2\left(-\frac{1}{\tau} \right) =\frac{1}{ \sqrt{\tau}} e^{-i\pi P_2(\tau)} Z_2(\tau), \qquad 	P_2(\tau) = \frac{1}{12} \left( \tau + \frac{1}{\tau}-3\right)\,,
\end{equation} 
We now recall how Cardy \cite{Cardy:1986ie} rederived this property from the Mellin representation of $Z_2(\tau)$: 
\begin{equation}\label{eq:Mellin-eta}
	\ln Z_2(\tau) = \frac{1}{2\pi i }\int_{\gamma-i\infty}^{\gamma+i \infty} ds (-2\pi i \tau)^{-s} \zeta(s+1) \Gamma(s) \zeta(s) \,.
\end{equation} 
In this case, the Hamiltonian zeta function is nothing but the Riemann zeta function $D(2,s)= \zeta(s)$. 
The modular transformation $\tau \to -1/\tau$ is translated to the $s \to -s$ transformation of Riemann zeta function. 
The modular image $Z_2\left(-\frac{1}{\tau} \right)$  is related to its original Mellin integral representation \eqref{eq:Mellin-eta} by the famous reflection formula of Riemann zeta function:
\begin{equation}\label{eq:zetasand1-s}
	\zeta(s) = \Gamma(1-s) 2^s \pi^{s-1} \sin \left(\frac{\pi s}{2} \right) \zeta(1-s) \,.
\end{equation}
Notice under the transformation $s \to -s$, the integration contour shifts $\gamma$ to $-\gamma$. 
We can now calculate the difference between the partition function at $\tau$ and $-1/\tau$ as a contour integral:
\begin{equation}
	\ln Z_2 \left( -\frac{1}{\tau}\right)-\ln Z_2(\tau)  = \frac{1}{2\pi i} \oint_{C} ds(-2\pi i \tau)^{-s} \Gamma(s) \zeta(s) \zeta(1+s)\,.
\end{equation}
The integration contour $C$ is is made up of the lines $[\gamma-i\infty, \gamma+i\infty]$, $[-\gamma-i\infty, -\gamma+i\infty]$ and infinity. 
By collecting the residues at the simple poles at $s=0,\pm 1$, one precisely reproduces the factor of automorphy \eqref{eq:modularity-eta}.
The asymptotics of the full partition function $Z_2(\tau,\bar{\tau})$ for $\tau,\bar{\tau} \to 0$ is given by 
\begin{equation}\label{eq:phase-2d-parti}
	Z_2(\tau,\bar{\tau}) \sim \frac{1}{\sqrt{\tau\bar{\tau}}}\exp \left[ \frac{\pi i (\tau+\bar{\tau})}{12} +\frac{\pi i}{12} \left( \frac{1}{\tau} +\frac{1}{\bar{\tau}}\right) - \frac{\pi i}{2} \right]\,.
\end{equation}
The inverse Legendre transformation leads to the Cardy formula for the density of states at large $\Delta\gg c$.

\subsection{\texorpdfstring{$d>2$}{d=2}}\label{ssec:mein-d>2}

Following the procedure outlined above, we rewrite the partition function \eqref{eq:scalar-partition-unrefined-d} of a $d$-dimensional free scalar as \cite{cardy1991operator,Kutasov:2000td,Dowker:2002ax,Gibbons:2006ij}: 
\begin{equation}
    \ln Z_d(\tau) = \frac{1}{2\pi i} \left(\oint_C+ \int_{-\gamma-i\infty}^{-\gamma+i \infty} \right) (-2\pi i \tau)^{-s} \Gamma(s) \zeta(s+1) D(d,s)ds \,,
\end{equation}
where the integral contour $C$ encloses all the simple poles of the integrand. 
Let us assume that $D(d,s)$ a finite number of simple poles at $s=\alpha_i$ with residue $A_i$, which holds for free field theory in various dimensions \cite{Melia:2020pzd}. 
One then finds
\begin{align}\label{eq:Meinardustheorem}
	\begin{split}
\ln Z_d (\tau)  &= \sum_i A_i \Gamma(\alpha_i) \zeta(\alpha_i+1) (-2\pi i \tau)^{-\alpha_i} -D(d,0) \ln(-2\pi i \tau) + D'(d,0) \\
&+ \frac{1}{2\pi i} \int_{-\gamma-i\infty}^{-\gamma+i \infty} ds (-2\pi i \tau)^{-s} \Gamma(s) \zeta(s+1) D(d,s)
	\end{split}
\end{align}
Here the first line contains the residues at the poles $s=\alpha_i$ and $s=0$ respectively. 
In $d=2$, the second line would become $\ln Z_2 (-1/\tau)$ after $s\to -s$ and using the reflection formula. 
For $d>2$, however, such a simplification does not occur. 
This reflects the fact that higher-dimensional partition functions are not modular, at least not in the same sense as in two dimensions.

Fortunately, for the $\tau \to 0$ asymptotics of the expression \eqref{eq:Meinardustheorem} is still useful. 
To compute the asymptotics, it was observed in \cite{Melia:2020pzd} that the ``bosonic'' Meinardus' theorem can be used \cite{elizalde1994zeta,andrews1998theory,Benvenuti:2006qr,Melia:2020pzd}.
This theorem implies that the second line in \eqref{eq:Meinardustheorem} is $\mathcal{O}(|\tau|^{C_0})$ when $\tau \to 0$ for some positive constant $C_0<1$.
To order $\mathcal{O}(\tau^0)$, one is thus left with the first line.
This line is straightforwardly computed through explicit evaluation of the Hamiltonian $\zeta$ function in \eqref{eq:ham-zeta} and its residues.

As a brief aside, it turns out that in $d=4$ the equation \eqref{eq:Meinardustheorem} can be interpreted as a modular property if one takes a $\tau$ derivative on both sides  \cite{cardy1991operator,Kutasov:2000td,Gibbons:2005vp}.
That is, consider the function defined by 
\begin{equation} \label{eq:modularity-Casimir-d4}
	E(\tau) = \frac{1}{2\pi i}\frac{\partial \ln Z(\tau)}{ \partial \tau}\,.
\end{equation}
Taking the $\tau$ derivative on both sides of \eqref{eq:Meinardustheorem}, one finds that the equation can be interpreted as a modular property of $E(\tau)$:
\begin{equation}\label{eq:Casimirenergy-transformation}
E(\tau) - \frac{B_4}{8}  = \frac{1}{\tau^4} \left[E\left( -\frac{1}{\tau}\right) - \frac{B_4}{8} \right] 
\end{equation}
Note that, up to a constant shift proportional to the fourth Bernoulli number $B_4=-\frac{1}{30}$, $E(\tau)$ transform as a weight $4$ modular form.
We will come back to this observation in Section \ref{sssec:4dscalar}.

Using \eqref{eq:Meinardustheorem}, it is straighforward to derive the asymptotics of the free scalar.
Explicitly, in $d=4,6$ dimensions one finds: 
\begin{eqnarray}\label{eq:asymptotics-4d-scalarpartition}
	\ln Z_4(\tau) &\sim& - \frac{ i\pi}{360\tau^3} + \zeta'(-2) +\mathcal{O}(\tau) \\
\label{eq:asymptotics-6d-scalarpartition}
	\ln Z_6(\tau) &\sim& \frac{i\pi }{15120 \tau^5}  +\frac{i \pi}{4320 \tau^{3}} + \frac{1}{12} \Big[ \zeta'(-4)- \zeta'(-2) \Big] +\mathcal{O}(\tau)
\end{eqnarray}
This method applies in general dimensions and can also be extended to free fermions \cite{Melia:2020pzd}.
The exact, all-order asymptotics was recently derived in an appendix of \cite{Benjamin:2023qsc} (based on the original work \cite{cardy1991operator}), making use of a $\zeta$-function regularization of the determinant of the $S^{d-1}\times S^1$ Laplacian \cite{elizalde1994zeta}.

In the following, we will argue that, in the case of the free scalar, there is in fact an interesting, albeit unconventional modular structure underlying the above results, despite the dimension being greater than two.
The relevant modular property provides an exact, closed form expression for the free scalar partition function which makes the high temperature asymptotics manifest.
Apart from recovering the above results, we also study implications for the fully refined partition function and the root-of-unity asymptotics $\tau\to 0^{+i}-\frac{n}{m}$.

\section{Exact asymptotics from generalized modularity}\label{sec:modularity-higher}

In this section, we will match the free scalar partition function in $D=2n$ with (a specialization of) certain multiple elliptic $\Gamma$ functions \cite{nishizawa2001elliptic}.
These functions are generalizations of the $q$-$\theta$ function and the elliptic $\Gamma$ function \cite{Felder_2000}, and satisfy interesting modular properties \cite{kurokawa2003multiple,narukawa2004modular}.
They naturally arise in the context of supersymmetric indices.
As we will see, the superconformal index of a $2D$-dimensional supersymmetric quantum field theory analytically continues the partition function of the $D$-dimensional, conformally coupled free scalar. 
This is an example of a more general observation in \cite{Nekrasov:2023xzm}, namely that the analytic continuation of parameters in a variety of physical quantities is naturally realized in supersymmetric quantum field theory.
After the identification, we show how the modular properties of multiple elliptic Gamma functions lead to a new closed form expression for the free scalar partition function which manifestly encodes the exact high temperature asymptotics.

Before starting, let us set up some notation following \cite{narukawa2004modular}.
We define chemical potentials $\tau_j$ and the associated fugacities $q_j$ by $q_j=e^{2\pi i \tau_j}$, where $0\le j\le r$.
Then we denote
\begin{align}
	\begin{split}
		\underline{q} = (q_0, \cdots, q_r), &\qquad \underline{\tau} = (\tau_0,\cdots,\tau_r) \\
		\underline{q}^-(j) = (q_0, \cdots, \check{q}_j, \cdots, q_r), & \qquad \underline{\tau}^-(j) = (\tau_0, \cdots, \check{\tau}_j, \cdots, \tau_r) \\
		\underline{q}[j] = (q_0, \cdots, q_j^{-1}, \cdots, q_r), & \qquad \underline{\tau}[j] = (\tau_0, \cdots, -\tau_j, \cdots, \tau_r) \\
		\underline{q}^{-1} = (q_0^{-1},  \cdots, q_r^{-1}), & \qquad -\underline{\tau}= (-\tau_0, \cdots, -\tau_r) 
	\end{split}
\end{align}
where $\check{q}_j$ means the $j$-th moduli is ruled out. 
We also define $|\underline{\tau}|=\tau_0+...+\tau_r$.
The generalized $q$-Pochhammer symbol is denoted by 
\begin{equation}\label{eq:def-q-Pochammer-r}
	(x;\underline{q})_{\infty}^{(r)} = \prod_{j_0,..,j_r=0}^\infty (1-x q_0^{j_0}q_1^{j_1}... q_r^{j_r})
\end{equation}
where $x= e^{2\pi i z}$ and which is defined for $\tau_j \in \mathbb{H}$ .
The generalized $q$-Pochhammer symbols satisfy 
\begin{equation}\label{eq:shift-refl}
    (x;\underline{q})_{\infty}^{(r)} = \frac{1}{(q_j^{-1}x; \underline{q}[j])^{(r)}_{\infty}}, \qquad (q_j x;\underline{q})^{(r)}_{\infty} = \frac{(x;\underline{q})_{\infty}^{(r)}}{(x;\underline{q}^- (j))_{\infty}^{(r-1)}}
\end{equation}
where the first property extends \eqref{eq:def-q-Pochammer-r} to $\tau_j \in \mathbb{C}\setminus\mathbb{R}$.
More generally:
\begin{equation}\label{eq:def-q-Pochammer-r-convergence}
	(x;\underline{q})_{\infty}^{(r)} = \left[ 
	\prod_{j_0,...,j_r=0}^\infty (1-x q_0^{-j_0-1}... q_{k-1}^{-j_{k-1}-1} q_k^{j_k}...q_r^{j_r})
	\right]^{(-1)^k}
\end{equation}
where we take $|q_0|,...,|q_{k-1}|>1$ and $|q_k|,...,|q_r|<1$.

\subsection{Modularity of the multiple elliptic Gamma function}\label{ssec:mod-mult-gamma}

The multiple elliptic Gamma function was first defined by Nishizawa in terms of the generalized $q$-Pochhammers defined above \cite{nishizawa2001elliptic}:\footnote{We collect various properties in Appendix \ref{app:spec-func}.}
\begin{equation}\label{eq:def-Gr}
	G_r(z|\underline{\tau}) = (q_0...q_r x^{-1};\underline{q})^{(r)}_{\infty} \cdot \left[(x;\underline{q})_{\infty}^{(r)} \right]^{(-1)^r}
\end{equation}
By using plethystic exponentiation (see \eqref{eq:Phythestic-exp}), these functions can be viewed as multi-particle partition functions whose single-particle partition function is given by 
\begin{equation}\label{eq:singleletter-Gr}
	Y_r(z|\underline{\tau}) = \frac{(-1)^{r+1} e^{2\pi iz}- e^{2\pi i(|\underline{\tau}|-z)}}{\prod_{i=0}^{r} (1-e^{2\pi i \tau_i})}
\end{equation}
As mentioned above, the function $G_r(z|\underline{\tau})$ naturally arises in the context of superconformal indices in dimension $D=2r+2$.
In particular, a single function $G_r$ represents the index of a free $D$-dimensional matter multiplet of minimal supersymmetry, as we will detail below.

The function $G_r(z|\underline{\tau})$ satisfies an interesting type of modular property \cite{narukawa2004modular} 
\begin{align}\label{eq:modularity-Gr}
	\begin{split}
		\prod_{k=1}^{r+2} G_{r} \left(\frac{\zeta}{\omega_k} \Big|\frac{\omega_1}{\omega_k},...,\frac{\check{\omega}_k}{\omega_k },...,\frac{\omega_{r+2}}{\omega_k}\right) =\exp \left[ -\frac{2\pi i}{(r+2)!} B_{r+2,r+2}(\zeta|\underline{\omega})\right]\,.
	\end{split}
\end{align}
Here, we use a homogeneous parametrization of the projective coordinates:
\begin{equation}
    (z|\underline{\tau})\equiv \left(\frac{\zeta}{\omega_{r+2}} \Big|\frac{\omega_1}{\omega_{r+2}},\ldots ,\frac{\omega_{r+1}}{\omega_{r+2}}\right)\,.
\end{equation}
We note that the equation is defined when the arguments of the $G_r$ functions are in $\mathbb{C}\setminus \mathbb{R}$, as follows from the definition \eqref{eq:def-Gr}.
That is, $\frac{\omega_j}{\omega_k}\notin \mathbb{R}$ for any $j,k$.
Furthermore, $B_{r,n}$ is the multiple Bernoulli polynomial defined by
\begin{equation}\label{eq:def-Bernoulli-poly}
	\frac{t^r e^{\zeta t}}{\prod_{i=1}^r (e^{\omega_i t}-1)} = \sum_{n=0}^\infty B_{rn}(\zeta|\underline{\omega})  \frac{t^n}{n!}\,.
\end{equation}
The relevant modular group for $G_{r}$ is $SL(r+2,\mathbb{Z})$.
For example, one may observe that the arguments of consecutive functions in \eqref{eq:modularity-Gr} are related by the $SL(r+2,\mathbb{Z})$ transformation which acts as a cyclic permutation.\footnote{For $r$ even, the cyclic permutation matrix has one entry with a $-$ sign so that the determinant condition is satisfied. This sign can be made manifest in \eqref{eq:modularity-Gr} through the extension property \eqref{eq:shift-refl}.}
The underlying modular structure was further explored in \cite{Tizzano:2014roa,Winding:2016wpw}.

More recently, based on early work on the elliptic Gamma function \cite{Felder_2000}, it has been proposed in \cite{Gadde:2020bov} that the modular property \eqref{eq:modularity-Gr} can be understood in terms of the group cohomology $H^r(SL(r+2,\mathbb{Z})\ltimes \mathbb{Z}^{r+2},N/M)$ with $N/M$ the space of meromorphic functions of $(z|\underline{\tau})$ modulo holomorphic, nowhere vanishing functions.
As before, the group $SL(r+2,\mathbb{Z})$ acts by matrix multiplication on the homogeneous parametrization $\omega_i$ of $\underline{\tau}$ while $ \mathbb{Z}^{r+2}$ shifts $\zeta$ with the periods $\omega_i$.
There appears to be a beautiful geometric interpretation of these properties in the context of superconformal indices, as suggested in \cite{Gadde:2020bov} and studied in detail in \cite{Jejjala:2022lrm} in four dimensions, which explains both the $SL(r+2,\mathbb{Z})$ modular group and the number of functions, $r+2$, involved in the property.
Let us briefly sketch this interpretation for $r=0,1,2,3$:
\begin{itemize}
    \item $r=0$: $G_0(z|\tau)^{-1}\equiv \theta(z;\tau)^{-1}$ arises as the elliptic genus of a 2d $\mathcal{N}=(0,2)$ chiral multiplet \cite{Kawai:1994np,Benini:2013nda}. 
    The modular group $SL(2,\mathbb{Z})$ reflects the large diffeomorphisms and $\mathbb{Z}^2$ the large gauge transformations associated with the $T^2$ geometric background.
    The modular property of $G_0(z;\tau)$ simply reflects the $S$-transformation of the $q$-$\theta$ function, which is well known to transform, up to a simple prefactor, as a weight zero, index one Jacobi form \cite{Felder_2000}. 
    \item $r=1$: $G_1(z|\tau_0,\tau_1)\equiv \Gamma(z;\tau_0,\tau_1)$ is the superconformal index of a 4d $\mN=1$ chiral multiplet \cite{Dolan:2008qi}. The group $SL(3,\mathbb{Z})\ltimes \mathbb{Z}^3$ is associated to a $T^3$ hidden in the $S^3\times S^1$ geometry. Indeed, one can view this geometry as a $T^2$ fibration over the toric manifold $\mathbb{CP}^1$ or, equivalently, as a $T^3$ fibered over an interval. 
    The modular property can be understood in terms of a bisection of $\mathbb{CP}^1$ into two disks, so that $S^3\times S^1$ is split as 
    \begin{equation}
        S^3\times S^1=D_2\times T^2 \overset{g}{\sqcup} D_2\times T^2\,.
    \end{equation}
    Here, $g\in SL(3,\mathbb{Z})$ exchanges the contractible cycle on the disk with a cycle on the torus, which is nothing but the genus one Heegaard splitting of $S^3$.
    The modular property of $G_1(z;\tau_0,\tau_1)$ reflects this geometric splitting, a phenomenon known as holomorphic block factorization \cite{Nieri:2015yia}.
    The three functions involved in this property can be associated to the three geometries, while the $SL(3,\mathbb{Z})$ transformations between consecutive arguments reflect the non-trivial identification $g$ and ambiguities in the ways the geometry can be split \cite{Jejjala:2022lrm}.\footnote{\label{fn:geom}
    One may wonder why the modular property treats all functions on an equal footing, while the geometry clearly does not. This is related to the fact that the one-loop determinants for a chiral multiplet on $S^3\times S^1$ and $D_2\times T^2$ have the same functional form \cite{Closset:2013sxa,Longhi:2019hdh}. This coincidence will be addressed in an upcoming work \cite{vanLeuven:2024}. }
    \item $r=2$: $G_2(z|\tau_0,\tau_1,\tau_2)^{-1}$ defines the 6d superconformal index of an $\mathcal{N}=(0,1)$ hypermultiplet \cite{Lockhart:2012vp,Imamura:2012efi,Spiridonov:2012de}. 
    Similar to the $r=1$ case, the group $SL(4,\mathbb{Z})\ltimes \mathbb{Z}^4$ is associated with a $T^4$ hidden in the $S^5\times S^1$ geometry. 
    Indeed, one can view this geometry as a $T^2$ fibration over the toric manifold $\mathbb{CP}^2$ or, equivalently, as a $T^4$ fibered over a triangle. 
    The modular property of $G_2(z|\tau_0,\tau_1,\tau_2)$ can be understood in terms of a trisection of $\mathbb{CP}^2$ in terms of three $D_2\times D_2$ components where each vertex of the triangle corresponds to the origin of a $D_2\times D_2$ component.
    Each component has an $S^3$ boundary and the intersection of the three boundaries is a $T^2$ located in the center of triangle.\footnote{See \cite{Gukov:2017zao} for a nice exposition on how to trisect general four-manifolds, including $\mathbb{CP}^2$.}
    In the full $S^5\times S^1$ geometry, the boundaries of the components have topology $S^3\times T^2$. 
    In order to specify how each component of the trisection is glued to the two others, the boundaries are bisected into two $D_2\times T^3$ geometries and each component of the bisection represents the interface between any two components of the trisection. 
    The bisections are thus labeled by three $SL(4,\mathbb{Z})$ transformations $g_{1,2,3}$, similar to the $r=1$ case.
    For consistency, $g_3$ is determined in terms of $g_{1,2}$.
    This explains how the trisection of $S^5\times S^1$ is labeled by effectively two $SL(4,\mathbb{Z})$ transformations.
    All in all, we write
    \begin{equation}
        S^5\times S^1=\bigsqcup_{i=1,2,3}\left(D_2\times D_2\times T^2\right)_{g_i} 
    \end{equation}
    The non-triviality of the $T^2$ fibration over $\mathbb{CP}^2$ is reflected in the transformations $g_{1,2,3}$, which again exchange contractible cycles in the disks with non-contractible cycles in the torus.
    The modular property of $G_2(z|\tau_0,\tau_1,\tau_2)$ reflects the fact that the 6d index respects the trisection \cite{Lockhart:2012vp,Qiu:2015rwp}.
    In analogy with the $r=1$ case, we expect the four functions involved in the modular property to be associated with these four geometries (see also footnote \ref{fn:geom}), and the consecutive $SL(4,\mathbb{Z})$ transformations on the arguments to follow from the identifications $g_{1,2,3}$ and ambiguities in the trisection of $S^5\times S^1$.
    However, we are not aware of a reference where this is fully worked out and leave the details to future work.
    \item $r=3$: it seems that $G_3(z|\tau_0,\tau_1,\tau_2,\tau_3)$ can be interpreted as an index in 8 dimensions. Some evidence for this can be found in \cite{Nekrasov:2017cih} and, through dimensional reduction to 7d and relations between the multiple Sine function $S_4$ and the multiple elliptic Gamma function $G_3$ (see Appendix \ref{appendix:multiplesine}), in \cite{Minahan:2015jta, Polydorou:2017jha,Polydorou:2020gxp, Minahan:2022rsx}. In this case, the group $SL(5,\mathbb{Z})$ arises by viewing $S^7\times S^1$ as a $T^2$ fibration over the toric manifold $\mathbb{CP}^3$ or, equivalently, as a $T^5$ fibered over a tetrahedron.
    We now quadrisect $\mathbb{CP}^3$ into four $D_2\times D_2\times D_2$ components such that each vertex of the tetrahedron corresponds to the origin of each component.
    Each component has an $S^5$ boundary and the components of the trisection of this $S^5$, as reviewed in previous bullet, correspond to the interfaces with the other three components.
    In the full geometry $S^7\times S^1$, these boundaries have topology $S^5\times T^2$.
    For each component $i=1,\ldots,4$, the trisection is therefore labeled by two $SL(5,\mathbb{Z})$ transformations $(g_1^{(i)},g_2^{(i)})$ . 
    All in all, we write for the quadrisection
    \begin{equation}
        S^7\times S^1=\bigsqcup_{i=1,2,3,4}\left(D_2\times D_2\times D_2\times T^2\right)_{(g_1^{(i)},g_2^{(i)})} 
    \end{equation}
    We are not aware of references studying this, but it is natural to conjecture that the 8d index respects this splitting and the five functions involved in the modular property can be associated to these five geometries, where we expect a similar comment as in footnote \ref{fn:geom} to apply.
\end{itemize}

Concluding, we see that the hierarchy of multiple elliptic Gamma functions $G_r$ and their modular properties fit quite naturally into expected factorization properties of $(2r+2)$-dimensional superconformal indices of the free minimally supersymmetric matter multiplet. 
The key to the modular properties in all these cases is that upon cutting open the manifolds, the geometric factors always contain a $T^2$ consisting of the time circle and the Hopf fiber of the odd-dimensional sphere.
Because of this, the splitting of the compact manifolds for fixed complex structure turns out to be ambiguous and can be labeled by elements in the $SL(2,\mathbb{Z})\subset SL(r+2,\mathbb{Z})$ subgroup corresponding to the large diffeomorphisms of these $T^2$s.
This was explored in detail in four dimensions \cite{Jejjala:2022lrm}, where the resulting modular structure was called modular factorization.
We expect that a very similar story extends to the other dimensions as well, along the lines of the above exposition.

\subsection{Identification with the free scalar partition function}\label{ssec:id-free-scalar}

Inspired by the observations of \cite{Nekrasov:2023xzm}, we now note that the partition function of the free scalar in \emph{even} dimensions $d=2D$ can be written in terms of $G_{d-1}$ by comparing the plethystic expressions in \eqref{eq:partitionfunction-anyD} and \eqref{eq:singleletter-Gr}.\footnote{The scalar partition function in odd $d=2D+1$ dimensional spacetime does not match with multiple elliptic Gamma functions.
In particular, the relative sign of the terms in the numerator of the plethystic exponent in \eqref{eq:partitionfunction-anyD} does not match the number of factors in the denominator of the multiple elliptic Gamma function $G_{2D}$. Nonetheless, the functions are still rather closely related and we will make some further comments on odd dimensions in Section \ref{sec:discussion}.}
For example,
\begin{itemize}
	\item In $d=2$, the partition function \eqref{eq:parti-2d} can be written in terms of the elliptic Gamma function: 
	\begin{equation}\label{eq:regulariza-2d-Gamma}
    Z_2(\tau,\bar{\tau}) = \lim_{z\to 0} (1-e^{2\pi i z}) \Gamma(z,\tau,\bar{\tau}) = Z_2(\tau)Z_2(\bar{\tau})
	\end{equation}
    This reflects the well-known fact that $\Gamma(z;\tau_0,\tau_1)$ has a simple pole at $z=0$ with a factorized residue $((q_0;q_0)_\infty(q_1;q_1)_\infty)^{-1}$.
    This divergence only occurs in $d=2$ and is related to the fact that free scalars in two dimensions are dimensionless.
	\item In $d=4$, we find a match with $G_3(z|\tau_0,\tau_1,\tau_2,\tau_3)$ through the identification:
    \begin{equation}\label{eq:moduli-4d}
	z=\tau,\quad \tau_0 = \tau+\theta_1, \quad \tau_1 = \tau+\theta_2, \quad \tau_2 = \tau-\theta_1, \quad \tau_3 = \tau-\theta_2\,,
    \end{equation}
    where we recall our convention \eqref{eq:beta-omega-tau-theta}.
    That is, we can write the 4d partition function as
    \begin{equation}\label{eq:4d-partitionfunction}
	Z_{4}(\tau,\theta_i) = G_3(\tau|  \tau+\theta_1, \tau+\theta_2 , \tau-\theta_1,  \tau-\theta_2 )\,.
    \end{equation} 
    \item In general $d=2D$, the partition function \eqref{eq:partitionfunction-anyD} can be written as 
    \begin{equation}\label{eq:Z2Dpartitionfunction}
    Z_{2D}(\tau,\theta_i)=G_{2D-1}((D-1)\tau| \underline{\tau})
    \end{equation}
    where 
    \begin{equation}
    \tau_{2i-2} = \tau-\theta_i,\quad \tau_{2i-1} = \tau+\theta_i, \qquad i=1,...,D
    \end{equation}
\end{itemize}
These identifications rely on two main facts: 1) the BPS operators contributing to superconformal indices in $2d$ dimensions contain only $d$ spacetime derivatives (see Section \ref{ssec:mod-mult-gamma} for references), matching the number of derivatives acting on the free scalar in $d$ dimensions, and 2) the contributions of BPS fermionic operators in the superconformal theory are weighed by $(-1)^F$. This relates closely to the constraint due to the equation of motion on the free scalar Hilbert space (see Section \ref{ssec:mellin}).

The identifications are somewhat reminiscent of the connection between Schur indices of 4d $\mathcal{N}=2$ SCFTs and characters of (non-unitary) VOAs \cite{Beem:2013sza}, and also of the connection between the 4d free, large $N$ Yang-Mills partition function in the confined phase with the chiral partition function of a 2d irrational CFT \cite{Basar:2015xda,Basar:2015asd}.
In particular, in the latter works a chemical potential $z$ is introduced in the 2d CFT which also has to be specialized for the matching to occur.
We will further comment on these works in Section \ref{sec:discussion}.

In the following, we use this identification and the modular properties of $G_r$ to present an exact, closed form expression for the partition function which is a function of the $S$-transformed variables $-\frac{1}{\tau_i}$.
We comment on a potential geometric interpretation in Section \ref{sec:discussion}.

\subsection{Unrefined partition function}\label{ssec:unrefined-partn}

In this section, we apply the modular property to calculate the asymptotics of the unrefined scalar partition function.
That is, we take $\theta_i=0$ so that all chemical potentials are equal: $\tau_i =\tau$.
We first set up notation and collect the relevant formulas.

For convenience of notation we define the unrefined multiple Gamma function as
\begin{equation}\label{eq:Def-Z4ztau}
    G_r(z;\tau) = G_r(z|\tau,\tau,...,\tau)
\end{equation}
Note that to obtain the $d$-dimensional scalar partition function, we have to further specialize $z$ as
\begin{equation}\label{eq:transunrefined}
Z_{d=r+1}(\tau)  \equiv G_r \left(\frac{r-1}{2}\tau;\tau \right)\,.
\end{equation}
We will sometimes refer to this specialization as the \emph{trans-unrefined} limit.
We also define the corresponding Bernoulli polynomial by
\begin{equation}\label{eq:def-Bernoulili}
    B_r(z;\tau) \equiv B_{rr} (z|\tau,\tau,...,\tau,1)\,
\end{equation}
where we have simply set $\omega_{r+2}=1$ to avoid writing the homogeneous variables.

When specializing the modular property \eqref{eq:modularity-Gr} for $\tau_i=\tau$, some care is required.
Indeed, this specialization implies (setting $\omega_{r+2}=1$):
\begin{equation}
    \omega_1=\ldots =\omega_{r+1}=\tau\,.
\end{equation}
Comparing with \eqref{eq:modularity-Gr}, we immediately note that all but one of the $G_r$ functions involved obtain real arguments, but recall that $G_r(z|\underline{\tau})$ is only defined when $\tau_i\in \mathbb{C}\setminus \mathbb{R}$.
However, it turns out that the resulting divergences cancel among the $G_r$ functions, and a well-defined non-trivial equation remains.\footnote{For the example of the elliptic Gamma function, this was shown in detail in Theorem 5.2 of \cite{Felder_2000} (see also \cite{narukawa2004modular}).}

To avoid dealing with these divergences, we present an alternative form of the modular property which is manifestly finite in the unrefined limit.
To get there, let us first recall Theorem 14 of Narukawa's work \cite{narukawa2004modular}.
This replaces the $G_r$ functions with an infinite product of normalized multiple Sine functions $S_r(z|\underline{\tau})$:
\begin{align}\label{eq:product-Gamma-Sine-2}
	\begin{split}
		G_r(z|\underline{\tau}) &= \exp\left[ -
		\frac{2\pi i}{(r+2)! } B_{r+2,r+2} (z|\underline{\tau},1)
		\right] \\
		&\times \prod_{k=0}^\infty \frac{S_{r+1}(z+k+1|\underline{\tau})^{(-1)^r} S_{r+1}(z-k|\underline{\tau})^{(-1)^r}}{\exp \{ \frac{i\pi}{(r+1)!} [ B_{r+1,r+1} (z+k+1|\underline{\tau}) - B_{r+1,r+1} (z-k|\underline{\tau})	]
			\}}
	\end{split}
\end{align}
where $S_r(z|\underline{\tau})$ is defined in Appendix \ref{appendix:multiplesine}.
This equation is defined for $\tau_j\in \mathbb{C}\setminus \mathbb{R}$ without any restriction on the \emph{ratios} of $\tau_j$.
Comparing with the conditions imposed below \eqref{eq:modularity-Gr}, we see that the domain of analyticity is enlarged.\footnote{This may be viewed as a generalization of how Faddeev's quantum dilogarithm extends a ratio of two ordinary $q$-Pochhammer symbols, naively defined for $\tau\in \mathbb{C}\setminus \mathbb{R}$, to $\mathbb{C}\setminus \mathbb{R}_{\leq 0}$ (see \cite{Faddeev:1994fw} or, \textit{e.g.}, the more recent in discussion in \cite{Dimofte:2009yn}). }
To recover the original form of the modular property from this expression, one uses the representation of the multiple Sine function in terms of generalized $q$-Pochhammer symbols \eqref{eq:explicit-Sr} and the definition of $G_r$ in \eqref{eq:def-Gr}.

There is, however, a remaining subtlety with the expression \eqref{eq:product-Gamma-Sine-2}: the infinite products of numerator and denominator is not separately defined; only the ratio is finite.
Let us denote this ratio by $\psi_{r}(z|\underline{\tau})$:\footnote{See Appendix \ref{appendix:multiplesine} for more details.}
\begin{equation}\label{eq:def-psifunction}
    \psi_r(z|\underline{\tau}) S_r(z|\underline{\tau}) = \exp \left[(-1)^{r-1}\frac{\pi i}{r!} B_{rr}(z|\underline{\tau}) \right]\,.
\end{equation}
In terms of this function, the modular property becomes 
\begin{align}\label{eq:product-Gamma-psi-2}
	\begin{split}
		G_r(z|\underline{\tau}) &= \exp\left[-
		\frac{2\pi i}{(r+2)! } B_{r+2,r+2} (z|\underline{\tau},1)
		\right]  \prod_{k=0}^\infty \frac{ \psi_{r+1}(z+k+1|\underline{\tau})^{(-1)^{r+1}}  } {
\psi_{r+1} \left(|\underline{\tau}|-z+k|\underline{\tau} \right) }
	\end{split}
\end{align}
where we used the reflection property\footnote{This can be derived from \eqref{eq:shift-refl} and \eqref{eq:Def-psir-function}.}
\begin{equation}
    \psi_r(z|\underline{\tau}) \psi_r(|\underline{\tau}|-z|\underline{\tau})^{(-1)^r} =1\,.
\end{equation}
We can now safely take the unrefined limit $\tau_i =\tau$.
We then use the homogeneity property \eqref{eq:homo-multiplesine} of $S_r(z|\underline{\tau})$ and \eqref{eq:def-psifunction} to write
\begin{equation}
\psi_r(z|\tau,...\tau) = \psi_r \left(\frac{z}{\tau} \Big|1,...,1 \right) \equiv \psi_r \left(\frac{z}{\tau}\right)\,.
\end{equation} 
This allows us to rewrite \eqref{eq:product-Gamma-psi-2} in the unrefined limit as 
\begin{align}\label{eq:Gr-mod-psir-S}
    \begin{split}
G(z;\tau) = \exp \left[
-\frac{2\pi i}{(r+2)!} B_{r+2,r+2} (z;\tau) 
\right] \prod_{k=0}^{\infty } \frac{\psi_{r+1} \left(\frac{z+k+1}{\tau}\right)^{(-1)^{r+1}}}{ \psi_{r+1} \big(\frac{(r+1)\tau-z+k}{\tau}\big)}\,.
    \end{split}
\end{align}
This is the modular property we will use in the remainder of this section.
We note that the specialization obscures the geometric interpretation of the modular property mentioned in Section \ref{ssec:mod-mult-gamma}.
It would be very interesting to find an alternative expression which, in the unrefined limit, reflects the underlying geometry more closely. 

Finally, we will make use of integral representations of these functions.
In particular, $S_r(z)\equiv S_r(z|1,\ldots,1)$ has an integral representation (see Appendix \ref{appendix:multiplesine})
\begin{equation}
    S_r(z) = \exp \left[ 
-\pi \int_{z_r}^z \left( 
\begin{array}{c}
     t-1  \\
      r-1
\end{array}
\right) \cot (\pi t) dt
    \right]\,.
\end{equation}
The $z_r$ indicate constants which are determined by $S_r(z_r)=1$. 
From here, we can work out the integral representation of $\psi_r(z)\equiv \psi_r(z|1,...,1 ) $ up to a constant $a_r$
\begin{equation}\label{eq:def-psi4function-2}
	\psi_r(z) = a_r\exp \left[  2\pi i
	\int_{-i\infty}^z \frac{1}{e^{2\pi i t}-1} \left(
	\begin{array}{c}
t-1		\\
	r-1	
	\end{array}
	\right) dt
	\right]\,.
\end{equation}
We can calculate $a_r = 1$ making use of the explicit values of $z_r$, which in turn can be determined from the values $S_r(1)$ 
\footnote{See also \eqref{eq:valueSr1} for the analytic expression for $S_r(1)$.} \cite{kurokawa2003multiple}. 
For example, we have
\begin{equation}
    S_2(1)=S_4(2)=S_6(3) =1\,.
\end{equation}
For other values of $r$, such values can also be worked out \cite{kurokawa2003multiple}.

\subsubsection{\texorpdfstring{$d=4$}{d=4}}\label{sssec:4dscalar}

We now calculate the all-order asymptotics of the unrefined 4d free scalar partition function, making use of the identification
\begin{equation}
    Z_{4}(\tau)=G_3(\tau;\tau)\,,
\end{equation}
and the modular property in the form \eqref{eq:Gr-mod-psir-S}:
\begin{equation}\label{eq:G3topsiproduct}
		G_3(z;\tau) = \exp \left[- \frac{2\pi i}{120} B_{5}(z+1;\tau)\right] \prod_{k=0}^\infty \frac{\psi_4 (\frac{z+k+1}{\tau})}{\psi_4(4-\frac{z-k}{\tau})}\,.
\end{equation}
We can find a completely explicit expression by carrying out the integral defining $\psi_4(z)$:
\begin{equation}\label{eq:def-psi4function-0}
	\psi_4(z) = \exp \left[  2\pi i
	\int_{-i\infty}^z \frac{(t-1)(t-2)(t-3)}{6(e^{2\pi i t}-1)} dt
	\right]
\end{equation}
which is defined for $\mathrm{Im}(z)<0$.
By expanding the geometric series and integrating terms, we find:
\begin{eqnarray}\nonumber
\psi_4(z) &= & \exp\left[\frac{(z-1)(z-2)(z-3)}{6} \ln(1-e^{-2\pi i z})  + \frac{i(3z^2-12z+11)}{12\pi} \text{Li}_2(e^{-2\pi i z})
\right. \\ \label{eq:explicitpsifunction}
&& +  \left.\frac{z-2}{4\pi^2}\text{Li}_3(e^{-2\pi i z})
-\frac{i}{8\pi^3} \text{Li}_4(e^{-2\pi i z})
\right]\,,
\end{eqnarray}
where $\mathrm{Li}_r(x)$ is the $r^{\text{th}}$ polylogarithm.
We now take the trans-unrefined limit $z=\tau$ to find:
\begin{equation}\label{eq:G3topsiproduct-2}
		G_3(\tau;\tau) = \exp \left[- \frac{2\pi i}{120} B_{5}(\tau+1;\tau)\right] \frac{1}{\psi_4(3)}\prod_{k=0}^\infty \frac{\psi_4 (\frac{\tau+k+1}{\tau})}{\psi_4(4-\frac{\tau-k-1}{\tau})}\,.
\end{equation}
The number $\psi_4(3)$ can be evaluated as:
\begin{equation}\label{eq:value-psi43}
		\psi_4(3) = \exp \left[  \frac{19 \pi i}{720} -\zeta'(-2) \right]\,.
\end{equation}
Together with the explicit expression \eqref{eq:explicitpsifunction}, the formula \eqref{eq:G3topsiproduct-2} provides a beautiful rewriting of the 4d scalar partition function.
In particular, this function manifests the exact high temperature asymptotics $\tau\to 0^{+i}$.
To see this, we first note that $\lim_{z \to -i\infty} \psi_4(z) \to 1+\mathcal{O}(e^{-2\pi i z})$.
Now, we have for Im$(\tau)>0$ that
\begin{align}\label{eq:validity-range}
	\begin{split}
&	\text{Im}\left( \frac{\tau+k+1}{\tau}\right) >0,  \quad (k \ge 0); \qquad \text{Im} \left(4-\frac{\tau-k-1}{\tau} \right)>0, \quad   (k \ge 0) \,.
	\end{split}
\end{align}
Therefore, in the limit  $\tau\to 0^{+i}$ the arguments of the $\psi_4(z)$ go to $-i\infty$, and we can combine \eqref{eq:G3topsiproduct-2} and \eqref{eq:value-psi43} to find: 
\begin{align}\label{eq:phaseofZ4}
	\begin{split}
	Z_4(\tau) &\sim  \exp \left[- \frac{2\pi i}{120} B_{5}(z+1;\tau)\right] \frac{1}{\psi_4(3)}  \left( 1+ \mathcal{O}(e^{-\frac{2\pi i }{\tau}}) \right) \\
	&=  \exp \left[
	-\frac{i\pi}{360 \tau^3} +\zeta'(-2) - \frac{i\pi \tau}{120}
	\right] \left( 1+ \mathcal{O}(e^{-\frac{2\pi i }{\tau}}) \right)\,.
	\end{split}
\end{align}
This result is consistent with Meinardus' theorem, as reviewed in Section \ref{ssec:mein-d>2}, but captures in addition the $\mathcal{O}(\tau)$ correction.
More importantly, our exact expression efficiently encodes all non-perturbative corrections as well.
To calculate those, we simply expand 
\begin{equation}
    \prod_{k=0}^\infty \frac{\psi_4 (\frac{\tau+k+1}{\tau})}{\psi_4(4-\frac{\tau-k-1}{\tau})}
\end{equation}
to any order $\mathcal{O}(e^{-\frac{2\pi i n }{\tau}})$ for $n\geq 0$, making use of the explicit expression for $\psi_4(z)$.
In particular, we note that the $k^{\text{th}}$ factor only contributes at order $\mathcal{O}(e^{-\frac{2\pi i (k+1) }{\tau}})$.
Equivalent results were recently obtained in Appendix C of \cite{Benjamin:2023qsc}, making use of the formalism described in Section \ref{ssec:mein-d>2}.
Our result can be viewed as a resummation of their exact asymptotics in terms of the special function $\psi_4(z)$.

The modular property \eqref{eq:G3topsiproduct-2} can be viewed as the integrated version of the more conventional modular property \eqref{eq:Casimirenergy-transformation} of $E(\tau)$, the $\tau$-derivative of $Z_{4}(\tau)$, as derived a long time ago by Cardy \cite{cardy1991operator} (see also \cite{Gibbons:2006ij}).
In particular, we can check that the $\tau$-derivative of our Bernoulli polynomial precisely reproduces their modular anomaly: 
\begin{equation}\label{eq:consistency-check-1}
	\partial_\tau \left[
	-\frac{2\pi i}{120} B_{5}(\tau+1;\tau)
	\right] =2\pi i \frac{B_4}{8}  \left(1-\frac{1}{\tau^4} \right)=E(\tau)-\frac{1}{\tau^4}E\left(-\frac{1}{\tau}\right) \,.
\end{equation}
An enlightening way to understand this modular property is to write $$E(\tau) =\sum_n \frac{d_n E_n}{e^{-2\pi i \tau E_n} -1}$$ and recall Ramanujan's formula \cite{Dowker:2002ax,Gibbons:2006ij}
\begin{equation}\label{eq:ramanujan}
    (\pi i \tau)^p \sum_n \frac{n^{2p-1}}{e^{2\pi i n \tau}-1} - \left(\frac{ \pi i}{\tau} \right)^p \sum_n \frac{n^{2p-1}}{e^{- \frac{2\pi i n}{\tau}}-1}  = \left[ 
    (\pi i\tau)^p - \left(\frac{ \pi i}{\tau} \right)^p 
    \right] \frac{B_{2p}}{4p}\,.
\end{equation}
The fact that the summand is close to the integrand in the integral representation of $\psi_r(z)$ explains why the derivative of partition function $\ln Z$ behaves as a modular form. 

The asymptotic growth of the density of states can be computed by inverse Legendre transformation followed by saddle point approximation, resulting in 
	\begin{equation}\label{eq:asymptoticgrowth-Z4}
		\rho(\Delta) \sim  \frac{1}{2\sqrt{2}\sqrt[8]{15} \left( \Delta+\frac{1}{240}\right)^{\frac{5}{8}}} \exp \left[\frac{4\pi}{3 \sqrt[4]{15}}  \left( \Delta+\frac{1}{240}\right)^{\frac{3}{4}} + \zeta'(-2) \right]\,.
\end{equation}
The difference between asymptotic growth \eqref{eq:asymptoticgrowth-Z4} and the prediction by Meinardus' theorem \cite{Melia:2020pzd} is the shift by $1/240$ in the conformal dimension. 
This reminds us of the Casimir energy $-c/24$ in two dimensional conformal field theory, which similarly appears as a shift in the conformal dimension in Cardy's formula \cite{Cardy:1986ie}.
To see this analogy more explicitly, we can consider $c$ free scalars. 
In this case, the asymptotics simply become
\begin{equation}
	Z(\tau) \equiv \exp \left[
	-\frac{ci\pi}{360 \tau^3} +c \zeta'(-2) - \frac{i\pi c\tau}{120}
	\right]\,. 
\end{equation}
The density of states, in turn, reads: 
\begin{equation}\label{eq:asymptoticgrowth-Z4-c}
		\rho \sim  \frac{c^{\frac{1}{8}}}{2\sqrt{2}\sqrt[8]{15} \left( \Delta+\frac{c}{240}\right)^{\frac{5}{8}}} \exp \left[\frac{4\pi}{3 \sqrt[4]{15}} c^{\frac{1}{4}} \left( \Delta+\frac{c}{240}\right)^{\frac{3}{4}} + c\zeta'(-2) \right]\,.
\end{equation}

\subsubsection{6d free scalar}

The free scalar theory in six dimensions can be analyzed similarly. 
In particular, the 6d partition function with angular potentials turned off can be written in terms of $G_5(z|\underline{\tau})$:
\begin{equation}\label{eq:transunrefined-d6}
	Z_6(\tau)= \exp \left[ \sum_{l=1}^\infty \frac{1}{l}
\frac{q^{ 2l } (1-q^{2l})}{(1-q^l)^6} 
\right]\equiv G_5(2\tau;\tau)\,.
\end{equation}
Specializing the modular property in the form \eqref{eq:product-Gamma-psi-2} for $\tau_i=\tau$ we find
\begin{align}\label{eq:modularity-equalmoduli-G5}
	\begin{split}
		G_5(z;\tau)
		&=\exp \left[- \frac{2\pi i}{7!} B_{7}(z+1,\tau)\right] \prod_{k=0}^\infty \frac{\psi_6 (\frac{z+k+1}{\tau})}{\psi_6(6-\frac{z-k}{\tau})}\,.
	\end{split}
\end{align}
From \eqref{eq:def-psi4function-2} we have the integral representation of $\psi_6(z)$:
\begin{equation}\label{eq:int-def-psi6}
	\psi_6(z) = \exp \left[2\pi i \int_{-i\infty}^z \frac{1}{e^{2\pi i t}-1}  \left( 
	\begin{array}{c}
		t-1	\\
		5	
	\end{array}
	\right)dt \right]\,.
\end{equation}
We now take $z=2\tau$ and calculate the $\tau\to 0^{+i}$ asymptotics in complete analogy with the 4d case. 
The asymptotics up to non-perturbative corrections is given by
\begin{equation}\label{eq:G5asymptotics}
	Z_6(\tau)  \sim \exp \left[ -\frac{2\pi i }{7!} B_{77}(2\tau+1,\tau)\right] \frac{1}{\psi_6(4)}\left( 1+ \mathcal{O}(e^{-\frac{2\pi i }{\tau}}) \right) \,.
\end{equation}
From the integral representation, $\psi_6(4)$ is evaluates to
\begin{equation}\label{eq:value-psi64}
	\psi_6(4) = \frac{120689i\pi}{60480} - 	\frac{1}{12} \Big[\zeta'(-4) -\zeta'(-2) \Big] \,.
\end{equation}
Therefore, the asymptotics of the 6d scalar partition function is 
\begin{equation}\label{eq:phaseZ6}
	\ln Z_6(\tau) \sim \frac{i\pi }{15120 \tau^5}  +\frac{i \pi}{4320 \tau^{3}} + \frac{1}{12} \Big[ \zeta'(-4)- \zeta'(-2) \Big] + \frac{31 i\pi \tau}{30240} +\mathcal{O}(e^{-\frac{2\pi i}{\tau}})\,.
\end{equation}
This result is again consistent with Meinardus' theorem \eqref{eq:asymptotics-6d-scalarpartition} but also produces the $\mathcal{O}(\tau)$ correction.
Furthermore, the non-perturbative corrections are also easily derived from our exact expression \eqref{eq:modularity-equalmoduli-G5}.
In particular, one first integrates \eqref{eq:int-def-psi6} to obtain an expression in terms of poly-logarithms similar to the 4d case, and then expand 
\begin{align}
	\begin{split}
		\prod_{k=0}^\infty \frac{\psi_6 (\frac{2\tau+k+1}{\tau})}{\psi_6(6-\frac{2\tau-k-1}{\tau})}\,.
	\end{split}
\end{align}
Again, this provides an exact, closed-form expression for the all order asymptotics, which resums the exact asymptotics obtained in \cite{Benjamin:2023qsc}.

\subsection{Refined partition function}\label{ssec:refined-partn}

We now extend our analysis to the 4d partition function including angular potentials.
This was identified with $G_3(z|\underline{\tau})$ in \eqref{eq:4d-partitionfunction}, which we recall for convenience:
\begin{equation}
	Z_{4}(\tau,\theta_i) = G_3(\tau|  \tau+\theta_1, \tau+\theta_2 , \tau-\theta_1,  \tau-\theta_2 )\,.
\end{equation} 
We write $\theta_i = \tau \Omega_i$ with $|\Omega_i|<1$ and find it convenient to again use the modular property in the form \eqref{eq:product-Gamma-psi-2}.
 Explicitly, the modular property reads
\begin{align}\label{eq:G3-Omegas}
	\begin{split}
	& G_3 \Big(\tau \Big|\tau(1+\Omega_1),\tau(1+\Omega_2),\tau(1-\Omega_1),\tau(1-\Omega_2) \Big) \\
		= & \exp\left[\frac{2\pi i}{120} B_{55} (\tau\big|\tau(1+\Omega_1),\tau(1+\Omega_2),\tau(1-\Omega_1),\tau(1-\Omega_2),-1 ) \right] \\
		&\times \prod_{k=0}^\infty \frac{\psi_4( \frac{\tau+k+1}{\tau}|1+\Omega_1,1+\Omega_2,1-\Omega_1,1-\Omega_2)}{\psi_4(3+\frac{k}{\tau}| 
			1+\Omega_1,1+\Omega_2,1-\Omega_1,1-\Omega_2	
			)}\,.
	\end{split}
\end{align}
Similar to the unrefined case, the $k=0$ term in the denominator contributes at $\mathcal{O}(\tau^0)$ while all other terms in the product contribute only at non-perturbative order $\mathcal{O}(e^{-\frac{2\pi i}{\tau}})$.
We notice, however, that in this case we require an integral representation of $\psi_4(z|\underline{\tau})$, which is considerably more complicated than the one for $\psi_4(z)$ \cite{narukawa2004modular}.
An alternative definition of $\psi_4(z|\underline{\tau})$, in terms of generalized $q$-Pochhammer symbols \eqref{eq:Def-psir-function}, does not apply since the arguments in this specialization are real.
For this reason, it is more difficult, although in principle possible, to calculate the non-perturbative corrections order by order in this case as compared with the unrefined case.

We can still easily calculate the $\tau\to 0$ asymptotics up to non-perturbative order.
For convenience, we define the $k=0$ term as
\begin{equation}
	\psi_4(3|1+\Omega_1,1+\Omega_2,1-\Omega_1,1-\Omega_2) = \exp [2\pi i\Psi(\Omega_1,\Omega_2)]
\end{equation}
for which we only have an implicit definition in terms of the integral representation of $\psi_4(z|\underline{\tau})$ \cite{narukawa2004modular}.
The asymptotics is then calculated as
\begin{align}\label{eq:phaseZ4-refined}
	\begin{split}
\ln Z_4(\tau, \tau\Omega_i) &= \frac{\pi i}{360\tau^3 (1-\Omega_1^2)(1-\Omega_2^2)} +\frac{ \pi i(\Omega_1^2 +\Omega_2^2)}{72(1-\Omega_1^2)(1-\Omega_2^2) \tau} - 2\pi i \Psi(\Omega_1,\Omega_2)\\
&+ \pi i \frac{3(\Omega_1^4+\Omega_2^4) +5 \Omega_1^2 \Omega_2^2-17 (\Omega_1^2+\Omega_2^2)-19}{720 (1-\Omega_1^2)(1-\Omega_2^2)} \\
&+ \pi i \frac{
	3(\Omega_1^4+\Omega_2^4) +5 \Omega_1^2 \Omega_2^2+3 (\Omega_1^2+\Omega_2^2)-3
}{360(1-\Omega_1^2)(1-\Omega_2^2)} \tau + \mathcal{O}\left(e^{-\frac{2\pi i }{\tau}}\right)
	\end{split}
\end{align}
where the rest of terms are non-perturbative in the $\tau\to0$ limit.

The asymptotics enables us to compute the asymptotic degeneracy of states for given energy \emph{and} angular momenta.
This was also studied in \cite{Benjamin:2023qsc}, but we include the analysis for completeness.

\subsubsection*{Small angular momenta}

Let's first consider the regime where the angular momenta are small compared to the energy but still large enough to do a saddle point approximation. 

We assume $\Delta ,J_1,J_2\gg 1$ but subject to
\begin{equation}
	1 \ll J_1,J_2 \ll \Delta\,.
\end{equation}
In this case, the chemical potentials related to angular momenta can be assumed to be much smaller than $1$. 
We can therefore consider the first term in \eqref{eq:phaseZ4-refined} and expand in terms of $\Omega_i$: 
\begin{equation}
	Q_5(\Omega_1,\Omega_2) \sim \frac{1}{720 \tau^3(1-\Omega_1^2)(1-\Omega_2^2)} = \frac{1}{720\tau^3} (1+\Omega_1^2+\Omega_2^2) +\mathcal{O}(\Omega^4)
\end{equation}
To calculate the density of state, we extremize:
\begin{equation}
	F_s(\tau,\Omega_1,\Omega_2) = -\frac{\pi i}{360 \tau^3} (1+\Omega_1^2+\Omega_2^2) -2\pi i \tau \Delta-2\pi i \tau \Omega_1 J_1-2\pi i \tau \Omega_2 J_2
\end{equation}
This results in the asymptotic density of states in the regime of interest:
\begin{equation}
	\ln \rho(\Delta,J_1,J_2) = \frac{2^{\frac{9}{4}} \pi\Delta^{\frac{3}{4}}}{\sqrt{15}} \left[ \left(
	\frac{x}{1-\sqrt{1-15 x}} \right)^{\frac{3}{4}}
	+\frac{1}{5} \left(\frac{1-\sqrt{1-15 x}}{x} \right)^{\frac{1}{4}}\right]
\end{equation}
where $x=\frac{J_1^2+J_2^2}{E^2}$. 
We can see when $x\to 0$, the \eqref{eq:asymptoticgrowth-Z4} is reproduced. 

\subsubsection*{Fast rotating regime}

We also consider the fast rotating limit for which $\Omega_i \to -1$. 
The limit $\Omega \to 1$ is similar due to the symmetry in \eqref{eq:G3-Omegas}. 
In this case, we parametrize $
	\Omega_i=-1-\omega_i
$. 
The entropy functional becomes 
\begin{equation}
	F_f(\tau,\omega_1,\omega_2)= - \frac{\pi i}{1440 \omega_1 \omega_2 \tau^3} -2\pi i (\Delta-J_1-J_2) +2\pi i \tau \omega_1 J_1 +2\pi i \tau \omega_2 J_2
\end{equation}
This functional is extremized by
\begin{equation}
	\omega_1= - \frac{\Delta-J_1-J_2}{J_1}, \qquad \omega_2= - \frac{\Delta-J_1-J_2}{J_2}\,,
\end{equation}
and the asymptotic density of states in this limit is computed to be
\begin{equation}
	\ln \rho(\Delta,J_1,J_2)= \frac{2\sqrt{2}}{\sqrt{3} \sqrt[4]{5}}\pi \sqrt[4]{(\Delta-J_1-J_2) J_1J_2}
\end{equation}
As we expect $\omega_i$ to be small, then $\Delta$ should be close to the unitarity bound:
\begin{equation}
	\Delta -J_1-J_2 \ge 0, \quad J_i>0\,.
\end{equation}

\section{Root-of-unity asymptotics} \label{sec:generalimodualr}

In Section \ref{sec:modularity-higher}, we derived explicit and exact expressions of free scalar partition functions in terms of the modular transformed variable $-\frac{1}{\tau}$.
To derive these expressions, we made use of the identification of the partition function with the multiple elliptic Gamma functions $G_r(z|\underline{\tau})$ and their modular properties.

In this section, we demonstrate how this connection can also be used to derive expressions in terms of a general modular transformed variable $\frac{k\tau+l}{m\tau+n}$.
The resulting expressions can then be used to calculate the ``root-of-unity'' asymptotics of the partition function, where $\tau\to-\frac{n}{m}+0^{+i}$.
Such limits are well-studied in the context of 2d CFT and play a prominent role in holography \cite{Dijkgraaf:2000fq,Maldacena:1998bw}.
The very recent work \cite{Benjamin:2024kdg} studies these limits in higher dimensional CFTs.\footnote{The root-of-unity asymptotics has also been of interest in the context of the 4d superconformal index, in particular in connection to AdS$_5$ black holes \cite{Cabo-Bizet:2019eaf,ArabiArdehali:2021nsx,Jejjala:2021hlt,Cabo-Bizet:2021plf,Choi:2023tiq}.}
As in Section \ref{sec:modularity-higher}, we will see that our method produces explicit, closed form expressions, which make use of the underlying modular structure, and make the asymptotics manifest.

\begin{figure}
	\centering
	\includegraphics[width=0.7\linewidth]{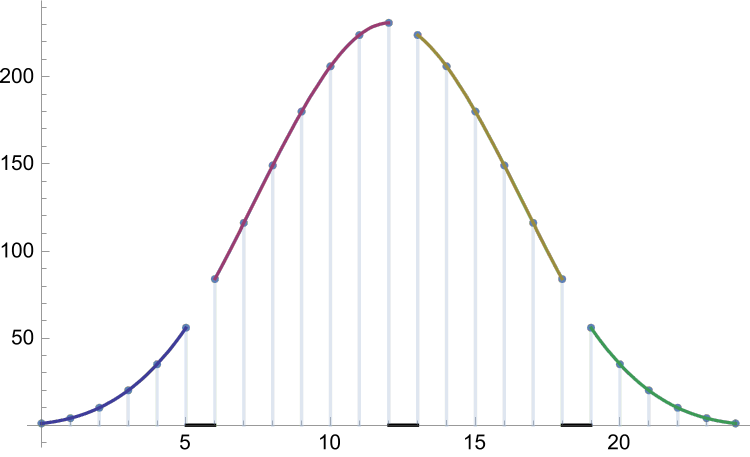}
	\caption{The piece-wise function $d_A(c)$ for $m=7$ as $0\le c \le 4m-4$. From left to right, the different colors represent difference pieces. 
 There are four disconnected pieces, labelled as I, II, III, IV. 
 Notice $d(c)$ are only defined on integer points. }
	\label{fig:m7}
\end{figure}

We start with a multiplication formula for $G_3(z|\underline{\tau})$: 
\begin{align}\label{eq:multiplication-G3}
\begin{split}
    &G_3(z|\tau_0,\tau_1,\tau_2,\tau_3)= \\
& \prod_{a_0=0}^{m-1} \prod_{a_1=0}^{m-1}\prod_{a_2=0}^{m-1}\prod_{a_3=0}^{m-1}
 G_3 \left(z+\sum_{i=0}^3 a_i \tau_i \Big| m\tau_0+n_0,m \tau_1+n_1, m\tau_2+n_2,m\tau_3+n_3 \right)
\end{split}
\end{align}
This formula is derived in Appendix \ref{appen-multiplication}.
We will only be interested in the unrefined limit with $\tau_i =\tau$.
In this case, the multiplication formula can be rewritten as
\begin{equation}\label{eq:mult-G3-equal-tau}
    G_3(z;\tau)= \prod_{c=0}^{4m-4} G_3(z+c\tau+N_c;m\tau+n)^{d(c)} 
\end{equation}
where $d(c)$ is an integer depending on $c$.
Its explicit piece-wise form is parametrized in terms of four functions $d_i(c)$ which are explicitly listed in Appendix \ref{appen-multiplication}.
We plot an example for $m=7$ in Figure \ref{fig:m7}.

We now obtain a formula in terms of the modular transformed variable by simply applying \eqref{eq:G3topsiproduct}, the modular property of $G_3(z;\tau)$, to each function on the right hand side of \eqref{eq:mult-G3-equal-tau}.
More specifically, each factor in \eqref{eq:mult-G3-equal-tau} is replaced by:
\begin{align}\label{eq:Z4-modularity-psi}
\begin{split}
		G_3(z+c\tau+N_c;m\tau+n)&= \exp\left[-
		\frac{  2\pi i}{120}B_{5} \left(z+c\tau+ N_c+1; m\tau+n \right) \right]  \\
  &\times \prod_{k=0}^\infty \frac{\psi_4( \frac{z+c\tau+N_c+k+1}{m\tau+n})}{\psi_4(4- \frac{z+c\tau+N_c-k}{m\tau+n})}    
\end{split}
\end{align}
The resulting formula provides us with an exact, albeit unwieldy expression for $G_3(z;\tau)$ in terms of a general modular transformed variable.
In particular, upon setting $z=\tau$, the formula encodes the all-order asymptotics of the unrefined 4d free scalar partition function in the limit $\tau\to-\frac{n}{m}+0^{+i}$.
Notice that the dependence on $N_c$ on the right hand side should cancel since the left hand side is periodic under $z\to z+1$.
We use this independence below to set $N_c$ such that the limit $\tau\to-\frac{n}{m}$ is most easily taken.

Before proceeding with the asymptotics, let us make some comments.
Compared to the analysis for $\tau\to -\frac{1}{\tau}$ in Section \ref{ssec:unrefined-partn}, in particular formula \eqref{eq:G3topsiproduct}, the number of factors in the resulting expression for $G_3(z;\tau)$ is much larger and in particular depends on $m$, the $SL(2,\mathbb{Z})$ parameter.
In principle, this does not pose a fundamental problem for, say,  the asymptotic analysis.
However, in practice, the expression becomes rather unwieldy to work with.
Furthermore, this formula obscures the already unconventional modular property in that the number of functions involved seems to depend on the $SL(2,\mathbb{Z})$ parameter.
We strongly suspect that, in fact, there exists a much simpler formula where all factors collapse into a single term.\footnote{See also \cite{Felder_2008,Tizzano:2014roa,Winding:2016wpw}.}
This suspicion is based on a similar phenomenon occurring for the elliptic Gamma function.
Very briefly, the derivation of the general modular property of the elliptic Gamma function in \cite{Jejjala:2022lrm} (Appendix D) involves the same strategy as above, but includes one further step.
Namely, just as above, one first uses the appropriate multiplication formula and applies the analogue of \eqref{eq:Z4-modularity-psi} to each factor.
However, in the final step, the multiplication formula is used again in the opposite direction to collapse the number of factors on the right hand side.
We leave this step in the present case to future work.

Let us now continue with the root-of-unity asymptotics.
Similar to Section \ref{ssec:unrefined-partn}, there are two types of contributions.
In terms of $\ln G_3(\tau;\tau)$, there are power law contributions coming from the Bernoulli polynomial and there is a constant contribution and non-perturbative contributions coming from the product over $\psi_4$ functions.
To take the limit, we now show that there is a unique choice of $N_c$ such that the $\psi_4$ functions do not diverge in the $\tau\to-\frac{n}{m}$ limit.
This choice turns out to be:
\begin{equation}\label{eq:valueNc}
	N_c = \bigg\lfloor\frac{(c+1)n}{m} \bigg\rfloor
\end{equation}
where $\lfloor \cdot\rfloor$ is the floor function.
To see this, just like in Section \ref{ssec:unrefined-partn}, we check that the arguments of all $\psi_4$ functions approach $-i\infty$ in the limit, or in exceptional cases, are constant. 
Setting $z=\tau$, the arguments of the $\psi_4$ functions in the numerator involve
\begin{equation}
	\tau+c\tau +N_c +k+1 = (c+1)\frac{(m\tau+n)-n}{m} +  \bigg\lfloor\frac{(c+1)n}{m}  \bigg\rfloor + k+1\,.
\end{equation} 
This is always positive for $\tau\to-\frac{n}{m}+0^{+i}$ so that
\begin{equation}
	\text{Im} \left(\frac{ z+c\tau+N_c+k+1}{m\tau+n} \right) \to -\infty
\end{equation}
and the functions $\psi_4 \left(\frac{ z+c\tau+N_c+k+1}{m\tau+n} \right) \to 1+\mathcal{O}(e^{-\frac{2\pi i}{m\tau+n}})$.
For the $\psi_4$ functions in the denominator, we similarly find 
\begin{equation}\label{eq:c+1criteria}
	4- \frac{\tau+c\tau+N_c-k}{m\tau+n} \to 4+ \frac{k+\frac{(c+1)n}{m}  - \lfloor\frac{(c+1)n}{m} \rfloor }{m\tau+n} - \frac{c+1}{m}\,.
\end{equation}
Again, the imaginary part generically goes to $-i\infty$ except when $ \lfloor\frac{(c+1)n}{m} \rfloor$ is an integer and $k=0$. 
These exceptions correspond to three values of $c$: 
\begin{equation}
	c=m-1, \quad 2m-1,\quad 3m-1\,.
\end{equation}
In these cases, the $k=0$ terms in the respective factors in \eqref{eq:mult-G3-equal-tau} contribute a constant listed in the following table:
\begin{table}[H]
	\centering
	\begin{tabular}{|c|c|c|}
		\hline
$c$ value		& $\psi_4$ value & Degeneracy \\
		\hline
$m-1$		& $\psi_4(3)$ & $d_{\text{II}}(m-1) $ \\
		\hline
$2m-1$		& $\psi_4(2)$ & $d_{\text{III}}(2m-1) $ \\
		\hline
$3m-1$		& $\psi_4(1)$ & $d_{\text{IV}}(3m-1) $ \\
		\hline
	\end{tabular}
\caption{Constant order contributions to the asymptotics of \eqref{eq:Z4-modularity-psi}. }
\label{table:constants-three}
\end{table}

Similar to the case of the elliptic Gamma function  \cite{Felder_2008,Jejjala:2021hlt,Jejjala:2022lrm}, the sum of all $B_5$ Bernoulli polynomials, coming from the replacement of \eqref{eq:Z4-modularity-psi} in  \eqref{eq:mult-G3-equal-tau}, for the choice of $N_c$ in \eqref{eq:valueNc} combine into a simple expression.
We will denote this sum by $Q_5^{(m,n)}(z;\tau)$ and find that it is equal to: 
\begin{equation}\label{eq:phaseQ5-0}
		Q_5^{(m,n)}(z;\tau) =\frac{1}{m} B_5(mz+n+1; m\tau+n) +C_5(m,n)
\end{equation}
where we refer to Appendix \ref{appen-multiplication} for more details.
We note that this holds even for $z\neq \tau$.
We have not determined a simple expression for the constant $C_5(m,n)$, but we note here that in the case of the elliptic Gamma function a similar constant can be expressed in terms of interesting number-theoretic quantities such as the Dedekind sum $s(n,m)$ and generalizations \cite{Jejjala:2022lrm}.  
An explicit formula for this constant is written in \eqref{eq:phaseQ5}.

Using the expression \eqref{eq:phaseQ5-0} and the constant contributions from  Table \ref{table:constants-three}, we find the leading order asymptotics for $\tau\to-\frac{n}{m}$ to be
\begin{align}\label{eq:phase-Z4-mn}
	\begin{split}
		G_3(\tau;\tau) 
	& \sim \exp \left[-2\pi i 	Q_5^{(m,n)}(\tau;\tau) \right] \exp\left[
	m^2\zeta'(-2) + \frac{i\pi (m^2-20)m}{720}\left(1+\mathcal{O}(e^{-\frac{2\pi i}{m\tau+n}})\right)
	\right]
	\end{split}
\end{align}
The real constant term in \eqref{eq:phase-Z4-mn} generalizes the prediction by Meinardus' theorem. 
We also note that the $\zeta'(-2)$ term has appeared before and was called the subleading topological entropy in \cite{Asorey:2012vp}. 
We thus find the root-of-unity asymptotics of the 4d free scalar partition function to be
\begin{align}
	\begin{split}
\ln Z_4(\tau) \sim& \frac{-i\pi}{360m(m\tau+n)^3}+ m^2 \zeta'(-2) -2\pi i C_5(m,n)\\
& +\frac{i\pi(m^2-1)(m^2-19)}{720m} - \frac{i\pi(m \tau+n)}{120m} \,.
	\end{split}
\end{align}
Pulling out $m$ in the denominator of the first terms, we obtain $1/m^4$ scaling at leading order consistent with \cite{Benjamin:2024kdg}.
From this equation, to leading order, we can evaluate the asymptotic density of states to be
\begin{equation}
	\rho(\Delta) \sim \exp \left(\frac{4\pi}{3 m \sqrt[4]{15}}\Delta^{\frac{3}{4}} \right)\,.
\end{equation}
The $1/m$ factor in the exponent indicates a $\mathbb{Z}_m$ quotient, which features explicitly in \cite{Benjamin:2024kdg}. 
The presence of such a factor was also noted in the context of superconformal indices at roots of unity \cite{Cabo-Bizet:2019eaf,ArabiArdehali:2021nsx,Jejjala:2021hlt,Aharony:2021zkr,Cabo-Bizet:2021plf}.

\section{Discussion and future work}\label{sec:discussion}

In this paper, we identified the multiple elliptic Gamma function $G_{r}$ with an analytic continuation, in temperature and angular potentials, of the thermal partition function of conformally coupled free scalars in \emph{even} dimensions $d=r+1$ on $S^{r}\times S^1$.

The natural habitat of the functions $G_{r}$, however, is in the context of superconformal indices of free minimally supersymmetric matter multiplets in dimension $2d=2r+2$, \textit{i.e.}, supersymmetric partition functions on $S^{2r+1}\times S^1$.
The modular properties satisfied by multiple elliptic Gamma functions have a beautiful geometric interpretation in terms of splitting properties of $S^{2r+1}\times S^1$ geometric backgrounds, as summarized in Section \ref{ssec:mod-mult-gamma}.
The torus consisting of the time circle and the Hopf fiber of the odd-dimensional spheres plays a crucial role.
As we have seen in Section \ref{ssec:id-free-scalar}, the specialization required to match the multiple elliptic Gamma function with the free scalar partition functions for real temperature and angular potentials obscures this geometric interpretation.
Nonetheless, one still finds an equation which expresses the free scalar partition function $Z_d(\tau)$ in terms of functions of a modular transformed variable $-\frac{1}{\tau}$, or more generally $\frac{k\tau+l}{m\tau+n}$.

A key question is now: does the geometric interpretation of the modular property in $d=2r+2$ teach us anything about the interpretation of the modular property of the free scalar in $d=r+1$?
Such an interpretation would be very valuable, with potential applications both to the underlying mathematics and also the extension to more general (non-supersymmetric) CFTs.
Naively, this seems tricky since almost everything is distinct: in $2d$ dimensions, we compute an index for a supersymmetric theory while in $d$ dimensions we compute a thermal partition function for a non-supersymmetric theory.\footnote{As we noted in Section \ref{ssec:id-free-scalar} and will further comment below, similar observations have however been made in different contexts. Sometimes, there is a good understanding of the identification, for example at the level of Hilbert spaces \cite{Beem:2013sza}.}
One way to make  progress may be to note that the variables $\underline{\tau}$ of the multiple elliptic Gamma function can be understood as complex structure moduli of the $S^{2r+1}\times S^1$ geometry.\footnote{See, \textit{e.g.}, \cite{Closset:2013vra,Assel:2014paa}, where the chemical potentials of the 4d superconformal index are matched with the complex structure moduli of a Hopf surface. A similar story applies to other dimensions as well.}
It seems possible that the specialization of these parameters, required to match the free scalar, singles out an $S^{r}\times S^1$ submanifold inside $S^{2r+1}\times S^1$.

As an example of what we mean by this in the case $r=3$, we note that $S^3\times S^1$ can be viewed as a submanifold of $S^7\times S^1$ through the $S^3\xhookrightarrow{} S^7\to S^4$ Hopf fibration of $S^7$.
It would be interesting to understand if the quadrisection of $S^7\times S^1$, as discussed in Section \ref{ssec:mod-mult-gamma}, can be effectively viewed as a bisection of $S^3\times S^1$.
Now, whereas the full modular property of $G_{3}(z;\underline{\tau})$ has a natural interpretation in terms of the quadrisection $S^7\times S^1$, it may be that the specialization is more naturally interpreted in terms of the bisection of $S^3\times S^1$.
More specifically, this would mean that the right hand side of the modular property:
\begin{equation}\label{eq:ex-result-disc}
		Z_4(\tau) = \exp \left[- \frac{2\pi i}{120} B_{5}(\tau+1;\tau)\right] \frac{1}{\psi_4(3)}\prod_{k=0}^\infty \frac{\psi_4 (\frac{\tau+k+1}{\tau})}{\psi_4(4-\frac{\tau-k-1}{\tau})}\,,
\end{equation}
could be interpreted in terms of the splitting of $S^3\times S^1$ into two $D_2\times T^2$ geometries and the associated concept of modular factorization \cite{Jejjala:2022lrm}.\footnote{In work in progress, we present a Hamiltonian picture of modular factorization in the context of 4d supersymmetric field theories. The emerging picture seems to suggest that this idea could be extended to (at least) free non-supersymmetric CFTs \cite{vanLeuven:2024}.}
To actually identify which terms on the right correspond to which $D_2\times T^2$ partition function, we see two ways forward.
Either, one should independently calculate the $D_2\times T^2$ partition functions and identify an appropriate product of two of such partition functions with \eqref{eq:ex-result-disc}.
Alternatively, one may attempt to derive an alternative version of \eqref{eq:ex-result-disc} which does not obscure the underlying geometry.
For example, starting off with the full modular property, one could try to take the specialization while carefully keeping track of the individual $G_3$ functions, which have a clear geometric origin.

We now mention some potential applications and generalizations of our results.
We first note that a similar relation has been found in the context of Schur indices of four-dimensional $\mathcal{N}=2$ SCFTs and characters of two-dimensional, non-unitary VOAs \cite{Beem:2013sza}.\footnote{See also \cite{Basar:2015xda,Basar:2015asd} for similar connections in the context of non-supersymmetric Yang-Mills theories.}
In particular, the $\mathcal{N}=2$ hypermultiplet is mapped onto a pair of symplectic bosons, which would be the analogue of our identifications of free chiral multiplets with free scalars, and the $\mathcal{N}=2$ vector multiplet is mapped onto $bc$ ghosts.
In their case, the matching extends to a matching between Hilbert spaces, or more precisely the $Q$-cohomology classes of Schur-operators in the SCFT and the VOA associated with the 2d CFT.
Central to our findings is that the appearance of a \emph{constraint}, namely the equation of motion, in the partition function of the free scalar becomes reinterpreted in terms of the higher-dimensional \emph{supersymmetric index} (see Section \ref{ssec:id-free-scalar}).
Non-unitary CFTs are well-known to implement (gauge) constraints in path integrals.
Even though the analogy is not perfect, the similarities are quite strong ann we believe it will be interesting to further explore.
For example, our observations could potentially shed light on the so-far somewhat mysterious role of non-unitarity in the Schur/VOA correspondence.
In the other direction, it would be very interesting if our match can be extended beyond the partition functions to a match between an appropriate $Q$-cohomology of the $2d$-dimensional chiral multiplet and local operators of the free scalar field theory.

As generalizations of our results, it is natural to ask what our findings could imply for more general, interacting CFTs.
Recently, the framework of thermal effective field theory was succesfully applied to the study of the high-temperature limit of general CFTs \cite{Benjamin:2023qsc,Allameh:2024qqp,Benjamin:2024kdg}.
The framework thus replaces the role of modularity in two-dimensional CFT. 
For the special case of the free scalar, it turns out that the framework fails but only in a minor way.
This is due to the fact that the free scalar, upon compactification on the thermal circle, has a gapless mode whose contribution is not captured by the formalism.
More specifically, the $\mathcal{O}(\beta^0)$ term in the high-temperature expansion (see \eqref{eq:phaseofZ4}) cannot be reproduced.
Another special, but not inconsistent, feature of the free scalar is that its expansion truncates after $\mathcal{O}(\beta)$, which is not expected for general CFTs.
On the other hand, non-perturbative corrections, as succinctly encoded in the full modular property \eqref{eq:ex-result-disc}, do take the form expected for general CFTs.
Our takeaway is twofold: having a simple, exact result for the free scalar could be helpful in understanding non-perturbative corrections for more general CFTs.
More speculatively, the fact that for the free scalar the modular property is able to capture the full thermal effective action, including the gapless contribution, and the thermal effective action is up to the Wilson coefficients completely universal, suggest perhaps a relation between at least an $\mathcal{O}(\beta)$ truncation of the thermal effective action for general CFTs and some notion of modularity.
If this were the case, it is conceivable that the modular property could relate the Wilson coefficients of the thermal effective action to conformal and/or 't Hooft anomalies of the CFT, which famously arise in modular properties of 2d CFTs, but also for indices of higher-dimensional SCFTs, see, \textit{e.g.}, \cite{DiPietro:2014bca,Gadde:2020bov}.

The modular properties observed for the free scalar may also extend to other, integrable quantum field theories. 
In fact, as noted in \cite{Felder_2000}, an early occurrence of the elliptic Gamma function was in the context of the free energy of the eight-vertex lattice model \cite{baxter1972partition}, see also \cite{Razamat:2013qfa,Nazzal:2023wtw} for more recent discussions in the context of elliptic integrable systems. 
We are not aware of connections between multiple elliptic Gamma functions $G_r(z|\underline{\tau})$ and integrable models. 
If such examples exist, they may provide additional examples where our generalized notion of modularity could find applications.

As immediate future directions, we would like to study:
(i) Scalars in odd dimension; (ii) Fermions in various dimensions;
(iii) Scalars in AdS, dS spaces \cite{Gibbons:2006ij}. 
The difficulty of constructing a modular property of the scalar partition function in odd dimensions ($d=3$) was already observed in \cite{cardy1991operator}.
Technically, this is due to the Hamiltonian zeta function $D(d,s)$ being a Hurwitz $\zeta$-function instead of a Riemann $\zeta$-function. 
An anti-periodic boundary condition along the time circle can ameliorate the issues \cite{cardy1991operator,Oshima:1992yy}. 
From our perspective, both free fermions in all dimensions and the free scalars in odd dimensions share the same obstacle: their partition functions are not of the form of the multiple elliptic Gamma function. 
However, the basic building blocks of their partition functions are generalized $q$-Pochhammer symbols, which also appear in the definition of the multiple elliptic Gamma functions. 
As such, these theories are not that far removed from our general framework either.
The original $q$-Pochhammer symbol $(x;q)_\infty$ has recently been understood as potentially the simplest example of a (holomorphic) quantum Jacobi form \cite{Garoufalidis:2022wij}.
It would be very interesting to generalize this notion to generalized $q$-Pochhammer symbols, and study the implications for the larger class of free CFTs as listed above along the lines of our work. 

A final interesting direction is to revisit the mysterious $SL(2,\mathbb{Z})$ duality between
the partition functions on two topologically distinct manifolds \cite{Shaghoulian:2016gol}:
\begin{equation}\label{eq:SL2Zmodualrity-shgou}
	\lim_{k,p \to \infty} Z \left[ S_{2\pi/k}^1 \times S^3/\mathbb{Z}_p\right] =	\lim_{k,p \to \infty} Z \left[ S_{2\pi/p}^1 \times S^3/\mathbb{Z}_k\right], \qquad k,p\in \mathbb{Z}^+
\end{equation}
This duality was checked for free field theory, supersymmetric theories and holographic CFTs.
It would be interesting if our perspective could shed some light on this proposal, perhaps also in the context of the recent work on twisted boundary conditions on the time circle \cite{Benjamin:2024kdg}.

\section*{Acknowledgement}
 
YL thanks 13th Joburg Workshop on String Theory and University of Witwatersrand for hospitality where this work was initiated. 
YL would also like to thank the organizers of the ``Frontier forum on Black holes and wormholes" 2024 in Thousands island lake county, the ICTP-AP UCAS and the String Math 2024 in ICTP for hospitality where this work is completed. 
Y.L. is supported by a Project Funded by the Priority Academic Program Development of Jiangsu Higher Education Institutions (PAPD) and by National Natural Science Foundation of China No.12305081 and No.12481540178.
SvL is in part supported by the DSI-NRF Centre of Excellence in Mathematical and Statistical Sciences (CoE-MaSS), South Africa, grant \#2022-59-Phy-Indices.
SvL would also like to thank the Tata Institute of Fundamental Research for hospitality, where part of this work was done.

\appendix

\section{Special functions}\label{app:spec-func}

\subsection{Multiple Sine function}\label{appendix:multiplesine}

The Barnes multiple zeta function was studied by \cite{barnes1904theory} as the generalization of the Hurwitz zeta function:\footnote{This is also called multiple zeta function, but the terminology ``multiple zeta function" is usually used as the following generalization of Riemann zeta function 
	\begin{equation}\label{eq:mzf}
		\zeta(s,t) = \sum_{m,n\ge 0}' \frac{1}{m^s n^t}
	\end{equation}
	We therefore refer to \eqref{eq:def-Barnes-MZF} as the \textbf{Barnes} multiple zeta function in order to distinguish from \eqref{eq:mzf}. 
} 
\begin{equation}\label{eq:def-Barnes-MZF}
	\zeta_r(s,z,\underline{\omega}) = \sum_{\vec{n} \ge 0} \frac{1}{(\vec{n} \cdot \underline{\omega}+ z)^s}
\end{equation}
The Barnes-MZ function \eqref{eq:def-Barnes-MZF} can be analytically continued to the complex plane $s\in \mathbb{C}$. 
The related multiple Gamma function is then defined as 
\begin{equation}
	\Gamma_r(z|\underline{\omega}) = \exp \left[
	-\frac{\partial}{\partial s}  \zeta_r(s,z,\underline{\omega}) \Big|_{s=0}
	\right]
\end{equation}
The multiple Sine functions $S_r(z|\underline{\omega})$ are defined as 
\begin{equation}\label{eq:def-multipleSine}
	S_r(z|\underline{\omega}) = 	\Gamma_r(z|\underline{\omega})^{-1} \Gamma_r(|\underline{\omega}|-z|\underline{\omega})^{(-1)^r} 
\end{equation}
The refined multiple sine functions have a few useful properties \cite{kurokawa2003multiple}. 
\begin{itemize}
	\item Shift property:
	\begin{equation}
		S_r(z+\omega_i|\underline{\omega}) = S_r(z| \underline{\omega}) S_{r-1}(z|\underline{\omega}^-(j))^{-1}
		\end{equation}
  where we use notation introduced at the beginning of Section \ref{sec:modularity-higher}.
	\item Multiplication formula:
	\begin{equation}\label{eq:multiplication-multiplesine}
		S_r(z|\underline{\omega}) = \prod_{ k_i=0}^{m-1} S_r \left(z+ \underline{k} \cdot \underline{\omega} |\,m \underline{\omega}\right)
	\end{equation}
	\item Homogeneity: for $c \in \mathbb{C}-\{ 0\}$
	\begin{equation}\label{eq:homo-multiplesine}
	S_r(cz|c\underline{\omega}) = 	S_r(z|\underline{\omega}) 
	\end{equation}
\item Reflection property:
\begin{equation}\label{eq:multiplesine-reflection}
	S_r(z| \underline{\omega}) S_r(|\underline{\omega}|-z|\underline{\omega})^{(-1)^r}=1
\end{equation}
\end{itemize}
The multiple sine function can be expressed in terms of the generalized $q$-Pochhammer symbol as \cite{narukawa2004modular}:
\begin{align}\label{eq:explicit-Sr}
	\begin{split}
		S_r(z|\underline{\tau}) &= \exp \left[ (-1)^r \frac{\pi i}{r!} B_{rr} (z|\underline{\tau})\right] \prod_{k=1}^r (x_k; \underline{q_k})_{\infty}^{(r-2)} \\
		&= \exp \left[ (-1)^{r-1} \frac{\pi i}{r!} B_{rr} (z|\underline{\tau})\right] \prod_{k=1}^r (x_k^{-1}; \underline{q_k^{-1}})_{\infty}^{(r-2)}
	\end{split}
\end{align}
provided that Im$(\tau_i/\tau_j) \neq 0$ and where we defined
\begin{align}\label{eq:xkqkdef-moudli}
	\begin{split}
		&x_k = e^{2\pi i \frac{z}{\tau_k}}, \quad q_{jk} = e^{2\pi i \frac{\tau_j}{\tau_k}}\,, \quad \underline{q_k}= (q_{1k},...,\check{q}_{kk},...,q_{rk})
	\end{split}
\end{align}
Furthermore, $B_{rr}(z|\underline{\tau})$ is degree $r$ polynomial in $z$ defined in \eqref{eq:def-Bernoulli-poly}.
It is sometimes convenient to introduce $\psi_r(z)$ functions which are defined as the multiple sine functions without the prefactor: 
\begin{equation}\label{eq:Def-psir-function}
	\psi_r(z|\underline{\tau})= \left[ \prod_{k=1}^{r}(x_k^{-1};\underline{q_k^{-1}})^{(r-2)}_{\infty} \right]^{-1}
\end{equation}

The two lines of \eqref{eq:explicit-Sr} are defined for $|q_j|<1$ and $|q_j|>1$ respectively. 
Due to the homogeneity condition, we can further simplify the multiple sine function by setting one of the elliptic moduli to $1$. 
For example, the $S_2(z|\tau,1)$ is  simply
\begin{equation}\label{eq:Definition-S2}
	S_2(z|\tau,1) = \exp \left[ \frac{\pi i}{2} B_{22}(z|\tau,1)\right] \prod_{j=0}^\infty \frac{1-e^{2\pi i (z+j\tau)}}{1-e^{2\pi i (\frac{z}{\tau} - \frac{j+1}{\tau})}} \,.
\end{equation}
This identity is essential for the holomorphic block factorization of 3d supersymmetric partition functions \cite{Nieri:2015yia}.

The multiple sine functions defined in \eqref{eq:def-multipleSine} are called \emph{normalized multiple sine function} in the literature \cite{kurokawa2003multiple}.
There is another type of multiple sine function called \emph{primitive multiple sine function} \cite{kurokawa2003multiple}, defined as 
\begin{align}\label{eq:def-primitive-multipleSine}
	\begin{split}
		\mathcal{S}_r(z) = \exp\left(\frac{z^{r-1}}{r-1} \right) \prod_{n=-\infty,\neq 0}^{+\infty} P_r \left(\frac{z}{n}\right)^{n^{r-1}}
	\end{split}
\end{align}
where 
\begin{equation}
	P_r (z) = (1-z) \exp \left(z+\frac{z^2}{2}+...+\frac{z^r}{r} \right)
\end{equation}
The paper \cite{kurokawa2003multiple} showed the $S_r(z)$ and $\mathcal{S}_r(z)$ are closely related. 
The relations is actually very transparent in the following representations 
\begin{align}\label{eq:differential-multiplesine}
	\begin{split}
		\frac{S_r'}{S_r}(z) &= (-1)^{r-1} \left( 
		\begin{array}{c}
			z-1	\\
			r-1	
		\end{array}
		\right) \pi \cot(\pi z) \\
		\frac{\mathcal{S}_r'}{\mathcal{S}_r}(z) &= z^{r-1} \pi \cot(\pi z)
	\end{split}
\end{align}
Given the differential form of multiple sine function \eqref{eq:differential-multiplesine}, the functions are completely fixed by an initial condition at special values. 
The Lemma 3.2 of \cite{kurokawa2003multiple} claims 
\begin{align}\label{eq:valueSr1}
	\begin{split}
		S_r(1) &= \exp\left(-2 \sum_{2\le k\le r-1,\text{even}} a(r,k) \zeta'(-k) \right) 
	\end{split}
\end{align}
where $a(r,k)$ are defined as 
\begin{equation}
	C_{X+r-2}^{r-1} =\sum_{k=1}^{r-1} a(r,k) X^k
\end{equation}
For concrete examples, we can use \eqref{eq:valueSr1} to determine a few special values such as
\begin{equation}
	S_4(2)=1, \quad  S_2(1)=1, \quad S_4(1) = e^{-\zeta'(-2)}
\end{equation}
These can then determine following integral representations of multiple sine functions 
\begin{align}\label{eq:examples-S2S4}
	\begin{split}
		& S_2(z) = \exp \left[
		- \pi \int_1^z (t-1)\cot(\pi t) dt	
		\right]\\
		&	S_4(z) = \exp \left[
		- \pi \int_2^z \frac{(t-1)(t-2)(t-3)}{6} \cot(\pi t) dt	
		\right]
	\end{split}
\end{align}
In order to derive the $S_2(z)$ from the refine expression \eqref{eq:Definition-S2}, we should carefully take the limit $\tau \to 1$.
It turns out we can apply the Theorem 5.2 of \cite{Felder_2000} as follows. 
\begin{align}\label{eq:S2-epsilon-expansion}
	\begin{split}
		\ln S_2 &= \frac{\pi i}{2} B_{22}(z|\tau,1) + \sum_{j=0}^\infty \ln (1-e^{2\pi i (z+j\tau)}) - \sum_{j=0}^\infty \ln (1-e^{2\pi i (\frac{z}{\tau} - \frac{j+1}{\tau}) }) \\
		&=  \frac{\pi i}{2} B_{22}(z|\tau,1)  - \sum_{l=1}^\infty \frac{1}{l}  \frac{e^{2\pi i l z}}{1-e^{2\pi i \tau l}} - \sum_{l=1}^\infty \frac{1}{l} \frac{e^{2\pi i l \frac{z-1}{\tau}}}{1-e^{-2\pi il/\tau}}
	\end{split}
\end{align}
By taking $\tau=1+\epsilon$ and expand the expression to the order $\epsilon^{(0)}$,  we can see the divergent piece in the limit $\epsilon \to 0$ will be cancelled and the finite piece will be non-trivial.
The result will exactly match with \eqref{eq:examples-S2S4} after explicit integration.

\subsection{Multiple elliptic Gamma function}\label{appendi-higherankd}

The multiple elliptic Gamma function is defined in the work by Nishizawa \cite{nishizawa2001elliptic} as a generalization of the elliptic Gamma function studied in \cite{Felder_2000}.
Given the the generalized q-Pochhammer symbol $(x;\underline{q})_{\infty}^{(r)} = \prod_{j_0,..,j_r=0}^\infty (1-x q_0^{j_0}q_1^{j_1}... q_r^{j_r})$, 
the definition of multiple elliptic Gamma function is 
\begin{equation}
	G_r(z|\underline{\tau}) = (q_0...q_r x^{-1};\underline{q})^{(r)}_{\infty} \cdot \left[(x;\underline{q})_{\infty}^{(r)} \right]^{(-1)^r}
\end{equation}
The known elliptic Gamma function $\Gamma(z,\tau,\sigma)$ studied in \cite{Felder_2000} is identically the $G_1(z|\tau,\sigma)$, while the $q$-$\theta$ function $\theta(z,\tau)$ is precisely the $G_0(z|\tau)$. 
For $r\in \mathbb{Z}$, the $G_r(z|\underline{\omega})$ construct a family of multiple elliptic Gamma functions. 
These functions share many similar properties with the elliptic Gamma functions. 
For example, the following properties are obeyed \cite{narukawa2004modular}:
\begin{align}\label{eq:properties-multipleGamma}
	\begin{split}
& G_r(z+1|\underline{\tau}) = G_r(z|\underline{\tau}),  \qquad G_r(z|\underline{\tau}) G(z-\tau_j\big|\underline{\tau}[j]) =1 \\
& G_r(-z|-\underline{\tau}) G(z|\underline{\tau}) =1, \quad G_r(z+\tau_j|\underline{\tau}) =G_r(z|\underline{\tau}) G_{r-1}(z| \underline{\tau}^-(j))\,.
	\end{split}
\end{align}

\section{Multiplication formula} \label{appen-multiplication}

The multiplication formula is a useful tool in relating elliptic Gamma functions with moduli $\tau$ to those with moduli $m \tau$ 
\cite{felder2002multiplication,Felder_2000}. 
It is also applied in \cite{Jejjala:2022lrm} to understand the holomorphic factorization of superconformal indices in the lens geometry $L(p,q)$. 
The multiplication formula of multiple elliptic Gamma function is based on the multiplication formula of generalized $q$-Pochammer symbols, which were defined in Appendix \ref{appendi-higherankd}.
The latter refers to the following simple rewriting:
\begin{equation}
    (x;\underline{q})_{\infty}^{(r)} =  \prod_{a_i=0}^{m-1}\prod_{j_i=0}^\infty (1-x q_0^{m j_0+a_0}... q_r^{m j_r+a_r})
    = \prod_{a_i=0}^{m-1} (x q_0^{a_0}...q_r^{a_r}, \underline{q^m})_\infty^{(r)}
\end{equation}
where $i=0,...,r$.
The $G_r(z|\underline{\tau})$ can then be shown to satisfy 
\begin{align}
	\begin{split}
 G_r(z|\underline{\tau} ) =\prod_{ \underline{a}=0 }^{m-1} G_r(z+ \underline{a} \cdot \underline{\tau}| m\underline{\tau}+\underline{n})
	\end{split}
\end{align}
where 
$
\underline{a} \cdot \underline{\tau} = \sum_{i=0}^{r} a_i \tau_i\,,m\underline{\tau}+\underline{n} = (m\tau_0+n_0\,,...,\,m \tau_r+n_r)
$.
A similar formula for the elliptic Gamma function $\Gamma(z,\tau,\sigma)$ was obtained in \cite{felder2002multiplication}. 

In this work, we are interested in unrefined limit of the multiplication formula, by taking $\tau_i=\tau\,, n_i=n$ for $G_3(z|\underline{\tau})$, as we studied in Section \ref{sec:generalimodualr}. 
The multiplication formula reduces to  
\begin{equation}\label{eq:product-G3-ai}
G_3(z\,;\tau) = \prod_{\underline{a}=0}^{m-1} G_3(z+\left(a_0+a_1+a_2+a_3\right) \tau\,; m\tau+n) 
%= \prod_{	c=0,\,\sum_ia_i =c}^{4m-4} G_3 (z+c\tau\,;m\tau+n)
\end{equation}
We then need to count for given integer $c$ subject to the condition 
\begin{equation}
	c=a_0+a_1+a_2+a_3, \qquad 0\le a_i \le m-1
\end{equation}
what is the degeneracy of $c$ for different values, considering the factors are independent of $a_i$.
The useful fact is that two variable summation can be decomposed as 
\begin{equation}
	\sum_{r=0}^{m-1} \sum_{s=0}^{m-1} = \sum_{c=0}^{m-2} \sum_{s=0}^c + \sum_{c=m-1}^{2m-2} \sum_{s=c+1-m}^{m-1}
	=\sum_{c=0}^{m-2} (c+1)+ \sum_{c=m-1}^{2m-2} (2m-1-c)
\end{equation}
We can use this formula to split the summation over $a_i$ into two pairwise groups. 
\begin{align}
	\begin{split}
&\sum_{a_0=0}^{m-1} \sum_{a_1=0}^{m-1} \sum_{a_2=0}^{m-1} \sum_{a_3=0}^{m-1} = \left(\sum_{a=0}^{m-2} \sum_{a_1=0}^a + \sum_{a=m-1}^{2m-2} \sum_{a_1=a+1-m}^{m-1} \right) \left(\sum_{b=0}^{m-2} \sum_{a_3=0}^b + \sum_{b=m-1}^{2m-2} \sum_{a_3=b+1-m}^{m-1} \right)
	\end{split}
	\end{align}
After summing over $a_1$ and $a_3$ we can then define $c=a+b$. Then
we can see four different regions of summation of $c$ will appear automatically 
\begin{align}
	\begin{split}
		& \text{I}:  \quad 0\le c \le m-2 \\
		& \text{II}:  \quad m-1\le c \le 2m-2 \\
		& \text{III}: \quad 2m-1 \le c\le 3m-3 \\
		& \text{IV}: \quad 3m-2 \le c\le 4m-4
	\end{split}
\end{align}
and degeneracy function $d(c)$ in each piece-wise domain will be
\begin{align}\label{eq:m4-fourregions}
	\begin{split}
		& d_{\text{I}}(c) = \frac{(c+1)(c+2)(c+3)}{6}, \\
		& d_{\text{II}}(c) = \frac{2m}{3}(m^2+11) +2 c (4 + c) m - 2 (2 + c) m^2
		-\frac{(c+1)(c+2)(c+3)}{2}, \\
		& d_{\text{III}}(c) =\frac{(c+1)(c+2)(c+3)}{2} - \frac{4}{3} (11 + 12 c + 3 c^2) m + 10 (2 + c) m^2 - \frac{22 m^3}{3} \\
		& d_{\text{IV}} (c)  =  \frac{(1 + c - 4 m) (2 + c - 4 m) (3 + c - 4 m)}{6} \,.
	\end{split}
\end{align}
For consistency checks, we can find 
\begin{equation}
	\sum_{c =0}^{m-2} d_{\text{I}}(c) +	\sum_{c =m-1}^{2m-2} d_{\text{II}}(c) +	\sum_{c =2m-1}^{3m-3} d_{\text{III}}(c) +	\sum_{c =3m-2}^{4m-4} d_{\text{IV}}(c) \equiv m^4
\end{equation}

We can then replace each triple elliptic Gamma function on RHS of \eqref{eq:product-G3-ai} by the corresponding phase polynomial $B_5$.
They will add up to give the phase polynomial of $G_3(z,\tau)$ as $\tau \to -n/m$:
\begin{align}\label{eq:phaseQ5}
	\begin{split}
		Q_5^{(m,n)}(z;\tau) &=\sum_{c=0}^{m-2}d_{\text{I}}(c)\, B_5 \left(z+c\tau+ \Big\lfloor \frac{(c+1)n}{m} \Big\rfloor+1;m\tau+n \right)  \\
		&+ \sum_{c=m-1}^{2m-2} d_{\text{II}}(c)\, B_5 \left(z+c\tau+ \Big\lfloor \frac{(c+1)n}{m} \Big\rfloor+1\,;m\tau+n \right) \\
		&+ \sum_{c=2m-1}^{3m-3}  d_{\text{III}}(c)\, B_5 \left(z+c\tau+ \Big\lfloor \frac{(c+1)n}{m} \Big\rfloor+1\,;m\tau+n \right) \\
		&+ \sum_{c=3m-2}^{4m-4}  d_{\text{IV}}(c)\,B_5 \left(z+c\tau+ \Big\lfloor \frac{(c+1)n}{m} \Big\rfloor+1\,;m\tau+n \right)\\
&=\frac{1}{m} B_5(mz+n+1;m\tau+n) +C_5(m,n)
	\end{split}
\end{align}

We can then use \eqref{eq:m4-fourregions} to compute the extra constants in \eqref{eq:phaseQ5}. 
Simply we can extract the constant term $K(m,c)$ in the $Q_5(z+c\tau+N_c,m\tau+n)$ and the ones in $\frac{1}{m} Q_5(mz+n ,m\tau+n)$. 
The computation directly gives
\begin{equation}
K(m,c) =	\frac{30 c^4 - 240 c^3 m + 660 c^2 m^2 - 720 c m^3 + 251 m^4}{12m^5} \left(m - 
	2 c n + 2 m \Big\lfloor \frac{(c+1)n}{m} \Big\rfloor \right)
\end{equation}
Then $C_5(m,n)$ in the phase \eqref{eq:phaseQ5} is precisely 
\begin{align}\label{eq:summand-constant}
	\begin{split}
C_5(m,n) &= \sum_{c=0}^{m-2} K(m,c) d_{\text{I}}(c) + \sum_{c=m-1}^{2m-2}  K(m,c) d_{\text{II}}(c) \\
&+ \sum_{c=2m-1}^{3m-3} K(m,c) d_{\text{III}}(c) + \sum_{c=3m-2}^{4m-4}  K(m,c) d_{\text{IV}}(c) - \frac{251(1+2n)}{12m}
	\end{split}
\end{align}
where the last rational function comes from the constant term in $\frac{1}{m}Q_5(mz+n;m \tau+n)$.

The summand in \eqref{eq:summand-constant} includes various structures 
\begin{equation}
	\mathcal{S}_{k,1}^{(i)}(n,m) =	\sum_{c=1}^{m-1} c^k  \left\lfloor\frac{(c-i)n}{m} \right\rfloor, \qquad k=4,5,6,7
\end{equation}
Some functions of these structures can be analytically computed by following the algorithm developed by Carlitz \cite{carlitz1975some}, for $i=0$.
For example, we can confirm by numerical checks
\begin{align}
	\begin{split}
		\mathcal{S}_{1,1}^{(0)} &= \frac{(m-1)(4m n -3m-2n)}{12}- m s(n,m)    \\
		\mathcal{S}_{2,1}^{(0)} &= \frac{m(m-1)(3m n -2m-3n+1)}{12}- m^2 s(n,m)
	\end{split}
\end{align}
where $s(n,m)$ is the Dedekind sum. 
However, it is not clear how to give an analytic formula for general $i$. 
In \eqref{eq:summand-constant}, there will be constants computed from the summand $	\mathcal{S}_{k,1}^{(i)}(n,m) $ for $i=0,1,2,3$.
The analytic form of these are not clear to our knowledge. 
But we can always do numerical checks for each given coprime pair $(m,n)$. 

\bibliographystyle{JHEP}
\bibliography{bib-bh}
  
 \end{document}